\documentclass[aps,twocolumn,floatfix]{revtex4}
\usepackage{times}
\usepackage{graphics}
\usepackage{epsfig,calc}

\usepackage{bm}

\usepackage{amsmath,amssymb}

\def\btt1{{\tt$\backslash$\string1}}%

\def\AmS{{\protect\the\textfont2
        A\kern-.1667em\lower.5ex\hbox{M}\kern-.125emS}}

\citeindextrue

\newcommand{\kt}{k_{\text{B}}T}
\newcommand{\ec}{\varepsilon_{\text{c}}}

\newcommand{\na}{n_\text{a}}
\newcommand{\gc}{g_\text{c}}
\newcommand{\eb}{\varepsilon_{\text{b}}}

\newcommand{\tmax}{\theta_{\text{m}}}
\newcommand{\pmax}{\phi_{\text{m}}}

\newcommand{\bvec}{\mathbf{b}}
\newcommand{\xvec}{\mathbf{x}}
\newcommand{\yvec}{\mathbf{y}}
\newcommand{\zvec}{\mathbf{z}}
\newcommand{\Rvec}{\mathbf{R}}
\newcommand{\rvec}{\mathbf{r}}

\begin{document}


\title{Controlling Viral Capsid Assembly with Templating}
\author{Michael F. Hagan}
\affiliation{Department of Physics, Brandeis University, Waltham, MA, 02454}
\date{\today}
\begin{abstract} We develop coarse-grained models that describe the dynamic encapsidation of functionalized nanoparticles by viral capsid proteins.  We find that some forms of cooperative interactions between protein subunits and nanoparticles can dramatically enhance rates and robustness of assembly, as compared to the spontaneous assembly of subunits into empty capsids.  For large core-subunit interactions, subunits adsorb onto core surfaces \emph{en masse} in a disordered manner, and then undergo a cooperative rearrangement into an ordered capsid structure.  These assembly pathways are unlike any identified for empty capsid formation.  Our models can be directly applied to recent experiments in which viral capsid proteins assemble around the functionalized inorganic nanoparticles [Sun et al., Proc. Natl. Acad. Sci (2007) 104, 1354].  In addition, we discuss broader implications for understanding the dynamic encapsidation of single-stranded genomic molecules during viral replication and for developing multicomponent nanostructured materials. 
\end{abstract}
\maketitle

\section{Introduction}

\label{sec:intro}
The assembly of simple building blocks into larger, ordered structures
is crucial for many biological processes and is enabling novel
nanostructured materials (e.g. (\cite{Glotzer2004,Whitesides2002b,Douglas2006,Valery2003,Yan2003,Strable2004,Blum2004}), which often draw inspiration and materials from
biology. The spontaneous assembly of viral proteins into empty capsids
(protein shells) has been the subject of elegant \emph{in vitro} experiments
(e.g. \cite{Johnson2005,Casini2004,Singh2003,Willits2003,Zlotnick2000,Klug1999,Zlotnick1996, Fox1994,Butler1978,Fraenkelconrat1955}) and insightful theoretical
works (e.g. \cite{Crick1956,Caspar1962,Berger1994,Zlotnick1994,Bruinsma2003, Zandi2004,Keef2005,Twarock2005,Hagan2006,Chen2007,Nguyen2007,Kegel2004,Hicks2006,vanderSchoot2007}).  The \emph{in vivo} replication of many viruses, however, involves
simultaneous assembly and encapsidation of the viral genome \cite{Dimmock2001}.  Likewise, many nanostructured materials require precise
spatial ordering of multiple, dissimilar components.  In this article,
we develop coarse-grained models for a
particular example of multicomponent assembly -- the assembly of viral
capsid proteins around a rigid spherical template.  Our models predict
that a template enables increased assembly rates and efficient assembly
over a wider range of parameters that control assembly driving forces,
as compared to the spontaneous assembly of empty capsids.  We find that template
properties can control assembly pathways, and our models predict a
novel mechanism that is unique to multicomponent assembly.

Our models are motivated by recent experiments in which Brome mosaic
virus (BMV) capsid proteins dynamically encapsidate functionalized
inorganic nanoparticle cores, creating unique biological and synthetic
composite structures called virus-like particles (VLPs) \cite{Sun2007,Dixit2006,Chen2005,Dragnea2003}.  By combining the
unparalleled self-assembly and targeting capabilities of viruses with
the functionalizability of nanoparticles, VLPs show promise as imaging
agents \cite{Soto2006,Sapsford2006,Boldogkoi2004,Dragnea2003}, diagnostic and therapeutic vectors \cite{Gupta2005,Garcea2004,Dietz2004}, and as subunits or templates for synthesis of
advanced nano-materials \cite{Chatterji2005,Falkner2005,Flynn2003,Douglas1998}.  Our models offer a framework with which to
interpret experimental results in order to design more efficient
templated assembly of nanomaterials, and a means to use this as a model
system with which to understand aspects of viral protein assembly
around nucleic acid cores.

Formation of the hollow shell geometry of a capsid poses a significant
challenge that requires anisotropic, directional interactions between
subunits.  Thus, in addition to their biomedical and technological
applications, studying viral capsids has revealed fundamental
principles of assembly.  Although specific assembly mechanisms are
poorly understood for most viruses, a general
mechanism has emerged for the spontaneous assembly of empty capsids \cite{Singh2003,Willits2003,Zlotnick2000,Klug1999,Zlotnick1996, Fox1994,Ceres2002,Endres2002,Hagan2006,Zhang2006,Zlotnick1994,Zlotnick2000,Zlotnick1999}.
Assembly occurs through a sequential addition process in which
individual subunits or larger intermediates \cite{Hagan2006,Zhang2006} bind to a growing capsid.  Assembly rates must be restrained
to avoid two forms of kinetic traps (long-lived metastable states): (a)
if new intermediates form too rapidly, the pool of free subunits
becomes depleted before most capsids finish assembling \cite{Ceres2002,Endres2002,Hagan2006,Zhang2006,Zlotnick1994,Zlotnick2000, Zlotnick1999}, (b) malformed structures result
when additional subunits bind more rapidly than strained bonds can
anneal within a partial capsid \cite{Sorger1986,Schwartz1998,Hagan2006,Nguyen2007}.  The formation of too many partial
capsids can be suppressed by a slow nucleation step \cite{Endres2002},
but avoidance of both sources of kinetic frustration requires
relatively weak subunit-subunit binding free energies \cite{Ceres2002,Endres2002, Hagan2006, Jack2007, Kegel2004,Nguyen2007}.  Theoretical work suggests that weak binding free energies
are a general requirement for successful assembly into an ordered low
free energy product; binding free energies that are large compared to
the thermal energy (kBT) prevent the system from `locally'
equilibrating between different metastable configurations during
assembly \cite{Jack2007, Whitesides2002}.

Although \emph{in vitro} studies of empty capsids provide a foundation for
understanding assembly, interaction of proteins with a central `core'
is crucial for the replication of many viruses in their native
environments, where capsid proteins must encapsidate the viral genome
during assembly \cite{Dimmock2001}.  There is no role for exogenous
species in the sequential assembly mechanism discussed above, but
\emph{in vitro} capsid assembly experiments in the presence of RNA demonstrate
different kinetics than capsid proteins alone, and suggest the presence
of protein-RNA intermediates \cite{Johnson2004}.

Prior theoretical and computational studies of multicomponent assembly
have examined the equilibrium behavior of polyelectrolyte encapsidation
\cite{Angelescu2006,vanderSchoot2005, Zhang2004,Hu2007b} and the
equilibrium configurations of colloids confined to convex surfaces
\cite{Chen2007b, Chen2007}.  A qualitative kinetic model has been
proposed to explain the formation of icosahedral symmetry in
encapsidated RNA \cite{Rudnick2005} and  Hu and Shklovskii \cite{Hu2007}
considered a model in which capsid proteins nonspecifically bind to
single-stranded RNA and slide on it towards an assembling capsid at one
end, which increases the rate of assembly.  While increased binding rates are one
possible feature of multi-component assembly, an interior core, such as a nucleic acid or nanoparticle, may also promote assembly by acting as a template that steers
assembly towards certain morphologies and as a heterogeneous nucleation
site that localizes capsid proteins in an environment favorable for
assembly.  These factors may generate assembly mechanisms that are
entirely different from the sequential mechanism considered in the
formation of empty capsids.  For instance, McPherson \cite{McPherson2005}
 proposed a qualitative model in which a large number of proteins
non-specifically bind to a nucleic acid molecule to form a structure
resembling a reverse micelle, and then reorient to form an ordered
capsid.  In this work we present a computational model for the
encapsidation of an interior core with no pre-assumed pathways.  For
some sets of system parameters, our simulations predict assembly
mechanisms consistent with McPherson's model.

In addition to the technological applications discussed above, solid nanoparticles offer a simplified, controllable experimental
system with which to test models for the effect of heterogeneous
nucleation and templating on assembly, and thus may lead to valuable
insights about viral assembly around nucleic acid cores as well as
elucidate the fundamental principles of multicomponent assembly.
Experiments show that capsid assembly around cores competes with
spontaneous assembly at subunit concentrations well above the threshold
concentration for empty capsid assembly (critical subunit
concentration, CSC) \cite{Sun2007}, and that assembly occurs in the
presence of cores below the CSC \cite{Dragnea2007}.  These results
point to the ability of nanoparticles to act as heterogeneous
nucleation agents.  In addition, nanoparticles promote formation of
capsid morphologies that are commensurate with nanoparticle sizes,
which suggests that cores can direct the final assembly product through
templating.  The time dependence of the mass averaged amount of
proteins on cores can be estimated by light scattering \cite{Dragnea2007}, but is not possible to characterize the extent to which these
proteins have assembled without static procedures, such as
crystallography or electron microscopy.  Kinetic models that relate
assembly pathways to dynamical observables such as light scattering are therefore
necessary to understand assembly mechanisms.

In this work, we present a computational model for
assembly around solid cores, with which we analyze kinetics and
assembly pathways as functions of the parameters that control the
driving forces of assembly, including subunit concentrations,
subunit-subunit binding energies, and surface adsorption free energies.
 At low adsorption free energies and/or low subunit concentrations,
assembly mechanisms resemble those seen for empty capsids, whereas
assembly pathways at high adsorption free energies and/or subunit
concentrations resemble the reverse micelle model.  We demonstrate that
the effect of cores on rates and assembly mechanisms can be understood
through simple and general scaling arguments.  

\section{Model}
\label{sec:model}

We consider a dilute solution of capsid subunits with a reduced
concentration $C_{ \text s } = \rho \sigma^3$, with $\rho$ the number
density and $\sigma$ the subunit diameter, and rigid cores with a
reduced concentration $C_{ \text C }$.  Subunits can spontaneously
assemble to form empty shells with well-defined structures of size $N$
subunits.  In addition, subunits interact favorably with cores and thus
adsorb to, and assemble on, core surfaces.  Complete assembly of
adsorbed subunits results in core encapsidation.  Our models are
motivated by the experiments described above in which viral capsid
proteins assemble on inorganic nanoparticles; thus, we begin by
adapting a model previously used to simulate the spontaneous assembly
of empty capsids \cite{Hagan2006}.  This computational model is
general, however, and could describe, for example, colloidal subunits
with directional interactions \cite{vanBlaaderen2006,Li2005,Cho2005}.
Likewise, the scaling arguments below are general enough to
describe many forms of simultaneous assembly and cargo encapsidation
for systems in which cargo degrees of freedom change slowly in
comparison to assembly timescales.

\subsection{Modeling empty capsid formation}
\label{sec:emptyCapsid}
\begin{figure} [bt]
\epsfig{file=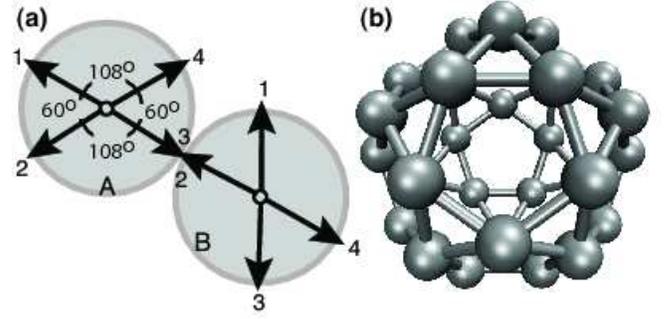,width= \linewidth}
\caption{\label{fig:one}
{\bf (a)} Geometry of subunits and attractive interactions for the computational model.  Bond vectors are depicted as arrows; bond vector 3 on subunit A and bond vector 2 on subunit B have an attractive interaction if they are nearly antiparallel and the secondary bond vectors, bond vector 2 on subunit A and bond vector 3 on subunit B, are nearly co-planar.  The angles between indicated bond vectors are specified in degrees.  {\bf (b)} The low free energy capsid geometry.  The center of each subunit roughly corresponds to a 2-fold axis of symmetry (at dimer interfaces) in a BMV T1 capsid (see Ref.~\onlinecite{Lucas2002} and the VIPER database \cite{Natarajan2005}).  Subunit sizes are reduced to aid visibility.
}
\end{figure}

  We imagine integrating over microscopic degrees of freedom as capsid
proteins fluctuate about their native states, to arrive at a pairwise
decomposable model in which subunits have spherically symmetric
excluded volumes and directionally specific, short ranged attractions
between complementary interfaces.  The lowest energy states in the
model correspond to separate `capsids', which consist of multiples of
60 monomers in a shell with icosahedral symmetry.  In this work we
model experiments in which BMV capsid proteins assemble around 6 nm
nanoparticles, for which only T1 capsid geometries are observed
\cite{Sun2007}.  Because the basic assembly unit of BMV is believed to
be a dimer and capsid proteins rapidly dimerize in solution
\cite{Speir1995}, our model subunit represents a protein dimer.  Our
energy minimum model capsid therefore is comprised of 60 monomers or $N
= 30$ dimer subunits arranged with icosahedral symmetry, as shown in
Fig.~\ref{fig:one}.

The locations of subunit interfaces are tracked by internal bond
vectors, $\bvec_i^{ ( \alpha ) }$, that are fixed rigidly within a
subunit frame of reference, with $\alpha \in \{ 1 , 2 , \dots , n_{
\text b} \}$ and $n_{ \text b }$  is the number of
interfaces on each subunit.  In this work there are $n_{ \text b } = 4$ bond vectors that can
be represented in Cartesian coordinates as $\bvec_i^{ ( \alpha ) } / b
= 0.5 k_x^{ \alpha } \hat { \xvec } + 0.809 k_y^{ \alpha } \hat { \yvec
} + 0.309 \hat { \zvec }$  with the bondlength $b=2^{-5/6}$ and $k_x = \{ 1 , - 1 , - 1 , 1 \}$  and
$k_y = \{ 1 , 1 , - 1 , - 1 \}$,  so that the angles between bond
vectors have the values indicated in Fig.~\ref{fig:one}a and the minimum
energy capsids have 30 subunits as shown in Fig.~\ref{fig:one}b. This
model results from merging pairs of monomeric subunits in the B5 capsid
model considered in Ref.~\onlinecite{Hagan2006} (see Fig.~1 in that
reference) and the resulting model capsid has the same connectivity as
a model considered by Endres et al. \cite{Endres2002} (see Fig 1B in
that reference).

The interaction between subunits $i$ and $j$ is
\begin{eqnarray}
u_{ij}&=&u_\text{rep}\left(|\mathbf{R}_i-\mathbf{R}_j|\right)+\sum_{\alpha \beta}^{\prime} u_\text{att}
\left(|\mathbf{r}_i^{(\alpha)}-\mathbf{r}_j^{(\beta)}|\right)\\
      & \times & s_{\alpha \beta}\left(\theta_{ij}^{(\alpha,\beta)},\theta_\text{m}\right)s_{\alpha,\beta}\left(\phi_{ij}^{(\gamma,\epsilon)},\phi_\text{m}\right)
\label{eq:uij}
\end{eqnarray}
where $\Rvec_i$ is the center of subunit $i$ and $\rvec_i^{ ( \alpha )
} \equiv \Rvec_i + \bvec_i^{ ( \alpha ) }$ is the position of
the interface represented by bond vector $\alpha$ on subunit $i$.  The
repulsive excluded volume interaction is
\begin{equation}
u_\text{rep}(R)=\Theta(r_\text{m}-R)[1+\mathcal{L}(R/\sigma)],
\label{eq:uwca}
\end{equation}
where we have defined a `Lennard-Jones' function
$\mathcal{L}(x)=4(x^{-12}-x^{-6}) $
and $\Theta ( x )$ is the step function with $r_m \equiv 2^{ 1
/ 6 } \sigma$ the maximum range of the repulsion.

The attractive interactions depend on the relative configurations and
alignments of complementary interfaces. To reflect this, the sum in Eq.~(\ref{eq:uij}) is marked with a prime to note that it runs over all pairs
of complementary interfaces, which in this work are all pairings of
bond vectors with an even and odd label, e.g. $( \alpha , \beta) = ( 1
, 4 )$ or $( 3 , 2 )$.  These pairs are denoted {\it primary}
interactions.  A favorable interaction between a pair of complementary
interfaces has three requirements (see Fig.~\ref{fig:one}a).  First, the interfaces should
closely approach each other, which is enforced by the distance
potential
\begin{equation}
u_\text{att}(r)=\varepsilon_\text{b}\Theta(r_\text{c}-\hat{r})\left[\mathcal{L}(\hat{r}/\sigma)-\mathcal{L}(r_\text{c}/\sigma)\right]
\label{eq:3}
\end{equation}
with $r_\text{c}=2.5$ the cutoff distance, $\eb$ the strength of the attractive interaction, and $\hat{r} \equiv r + r_m$ is a shifted distance so that the minimum of the potential occurs at $r=0$.  The second requirement for a favorable interaction is that primary bond vectors,$\bvec_i^{(\alpha)}$ and $\bvec_j^{(\beta)}$ are aligned antiparallel, which is enforced by the first factor on the second line of Eq. (\ref{eq:3}) with
\begin{equation}
\cos \left(\theta_{ij}^{(\alpha,\beta)}\right)=-\bvec_i^{(\alpha)} \cdot \bvec_j^{(\beta)}/b^2
\label{eq:4}.
\end{equation}
The tolerance of the potential to angular fluctuations is controlled by the specificity function $s_{\alpha\beta}$:
\begin{equation}
\label{eq:sab}
s_{\alpha\beta}(\psi,\psi_\text{m})\equiv\frac{1}{2}\Theta(\psi_\text{m}-\psi)[\cos(\pi\psi/\psi_\text{m})+1]
\end{equation}
where $\psi_\text{max}$ is the maximum angle deviation, which controls the angular specificity of the attractive interactions.  In this work, we vary the primary bond angle tolerance, $\tmax$, from 0.25 to 2.5 radians. 

The third requirement for an attractive interaction is that two {\it secondary} bond vectors, which are not involved in the primary interaction, are coplanar.  This requirement enforces angular specificity in the direction azimuthal to the primary bond vectors.  A specific pair of secondary bond vectors $(\gamma,\epsilon)$ is associated with each primary pair $(\alpha,\beta)$, and the second angular factor in Eq. (\ref{eq:uij}) favors the alignment of the normals to two planes.  The first plane is defined by the inter-subunit vector $\Rvec_{ij}\equiv \Rvec_i-\Rvec_j$ and the first member of the secondary pair, $\bvec_i^{(\gamma)}$; the second plane is defined by $\Rvec_{ij}$ and the second member of the same secondary pair, $\bvec_j^{(\epsilon)}$.  Denoting these normals by $\mathbf{n}_i^\gamma \equiv \bvec_i^{(\gamma)} \times \Rvec_{ij} $ and $\mathbf{n}_j^\epsilon \equiv \bvec_j^{(\epsilon)} \times \Rvec_{ij}$, the dihedral angle $\phi$ in Eq.~(\ref{eq:uij}) is determined from
\begin{equation}
\cos(\phi_{ij}^{\alpha,\beta})=\hat{\mathbf{n}}_{i}^{(\gamma)} \cdotp \hat{\mathbf{n}}_{j}^{(\epsilon)}
\label{eq:6}
\end{equation}
with $\hat{\mathbf{n}}$ the unit vector $\mathbf{n}/|n|$.

  Subunit positions and orientations are propagated according to overdamped Brownian dynamics, with the unit of time $t_0=\sigma^2/48D$, where $D$ is the subunit diffusion coefficient.  All energies are measured in units of the thermal energy $\kt$.
  In this work, secondary pairs are specified as the inverse of the primary pairs: $(\gamma,\epsilon)=(\beta,\alpha)$, and the dihedral specificity parameter $\pmax=\pi$ radians throughout. 
We vary the subunit concentration over the range $C_\text{S} \in  [2 \times 10^{-3}, 4 \times 10^{-2}]$.  If we choose the diameter of the dimer subunit to be $\sigma=5$ nm, these concentrations correspond to 27--540 $\mu$M.
  
  We note that Nguyen and coworkers \cite{Nguyen2007} recently developed a model for capsids subunits in which subunit excluded volumes have a roughly trapezoidal shape.  Interestingly, assembly kinetics predicted by their model are qualitatively similar to those of the model described above \cite{Hagan2006}, except that they find insertion of the final subunit (to form a complete capsid) is rate limiting, and uphill in free energy for many sets of parameters.  While insertion of the final subunit is also impeded by excluded volume for the present model, we find that once completed, capsids are stable and dissociation of a subunit is slow compared to assembly timescales.  This result seems consistent with experimental observations that subunit exchange between completed P22 capsids and free subunits is characterized by long timescales (days) compared to those for assembly (minutes) \cite{Parent2007, Prevelige1993}. 

\subsection{Modeling nanoparticle encapsidation}
We modify the empty capsid model by introducing a nanoparticle, or a rigid sphere, with radius $R_\text{C}$ at a fixed position at the center of the simulation cell ${\mathbf C}$.  In addition to the pairwise interactions between subunits described above, subunits interact with the nanoparticle via excluded volume interactions and attractive interactions.  The potential energy of interaction  $u_{\text{C}}(|\mathbf{R}-\mathbf{C}|) $ between a subunit at position $R$ and the nanoparticle is a spherically symmetric shifted Lennard-Jones potential
\begin{equation}
\label{eq:7}
u_{\text{c}}(r)=\varepsilon_{\text{c}}\Theta(r_{\text{c}}-\hat{r})\left[\mathcal{L}(\hat{r}/\sigma)-\mathcal{L}(r_c/\sigma)\right]
\end{equation}
where $r_\text{c}=2.5$ is the cutoff distance.  The strength of the attractive interaction is dictated by $\varepsilon_{\text{c}}$ and $\hat{r}=r-(R_{\text{C}}-0.5) $ is a shifted distance so that the attraction has its minimum value, $-\ec-u_{\text{c}}(r_{\text{c}}) $ , when the center of the subunit and the surface of the nanoparticle are separated by  $(2^{1/6}-0.5) \sigma $, and maintains the short range nature of the interactions considered in Ref. \cite{Hagan2006}.  This potential mimics core-subunit interactions that do not favor particular subunit orientations; we consider electrostatic interactions that depend on subunit orientations in a future work.  The core-subunit interaction free energy, $g_\text{c}\equiv \ec -T s_\text{ad}$, includes an entropy penalty, $s_\text{ad}$, for frozen degrees of freedom in the direction normal to the surface.  We determine $s_\text{ad}$ by calculating the partition function of an adsorbed subunit according to Eq.~(\ref{eq:7}); the result varies weakly with the adsorption energy: $1<s_\text{ad}/k_\text{B} <2.4$  for $3\le \ec \le 12$.  

\emph{Subunit reservoir}.  We represent a nanoparticle immersed in bulk solution without explicitly simulating thousands of subunits by coupling dynamical simulations to a reservoir of subunits at constant chemical potential.  We divide the simulation box into a "main" region centered around the core, where ordinary dynamics are performed, and an outer "bath" region where, in addition to ordinary dynamics, subunits are inserted or deleted in grand canonical Monte Carlo moves \cite{Frenkel2002}.  The main region is chosen to have a side length of  $L_\text{m}=15 \sigma$, which is large enough that subunits cannot simultaneously interact with the nanoparticle and a subunit in the bath area, while the bath has a total width of $L_\text{b}=6\sigma $.  Insertions and deletions are attempted with a frequency consistent with the diffusion limited rate for a spherical volume with a diameter of $L_\text{m}+L_\text{b} $.  As assembly proceeds, the concentration of free subunits is depleted and the bath chemical potential should be updated self-consistently.  In this work, we consider assembly around a single nanoparticle in infinite dilution (i.e. $C_\text{C} = 0$) at system parameters for which little or no spontaneous assembly occurs away from cores, as we will see in the next section. Hence, the chemical potential remains constant.

\section{Results}
\label{sec:results}
We have simulated assembly dynamics over ranges of subunit concentrations $C_\text{S}$, binding energies $\eb$, subunit specificity parameters $\tmax$, and surface attraction strengths $\ec$. All simulations use $\pmax =\pi$  and $R_\text{C}=1.2$.

\subsection{The kinetics of core-controlled assembly}
In this section we present simple scaling arguments for the effect of core-subunit interactions on the kinetics of assembly and illustrate scaling in simulation assembly trajectories.  We first concentrate on the average time to form a capsid, starting from unassembled subunits.  As shown by Zlotnick and coworkers \cite{Zlotnick1999,Endres2002}, the assembly of empty capsids can often be broken into nucleation and elongation phases.  We show elsewhere\cite{Kamber2007} that the average timescales of these phases for an individual capsid can be described by $\tau=\tau_{\text{nuc}}+\tau_{\text{elong}}$, with $\tau_{\text{nuc}}^{-1}\propto f c_1^{n_{\text{nuc}}} $  and $\tau_{\text{elong}}^{-1}\approx c_1 f /(N-n_{\text{nuc}}) $, where $f$ is the subunit-subunit binding rate constant, $c_1$ is the concentration of free subunits, $n_\text{nuc}$ is the number of subunits in the nucleus, and we assume $c_1$ remains roughly constant during the assembly of an individual capsid.  Because elongation requires $N-n_\text{nuc}$ assembly events, it introduces a minimum timescale for the overall assembly process, which is primarily responsible for the lag time in assembly kinetics reported in experiments \cite{Casini2004,Zlotnick2000,Zlotnick1999}, theory \cite{Endres2002,Zlotnick1994}, and simulations \cite{Hagan2006,Nguyen2007,Wilber2007}, and results in a distribution of assembly times for an individual capsid that cannot be fit with a sum of pure exponential functions\cite{Kamber2007}.  The observed assembly rate constant, $f$, can be considered an average quantity, since computational models \cite{Hagan2006,Nguyen2007} suggest that it varies for different intermediates and decreases due to excluded volume constraints as assembly nears completion.  In one model \cite{Nguyen2007}, insertion of the final subunit is slow compared to the rest of elongation and thus introduces a third timescale.  Zandi, van der Schoot, and coworkers use continuum theory approaches to analyze nucleation\cite{Zandi2006} and capsid formation rates at long times\cite{vanderSchoot2007}.  Their finding that the total rate of capsid formation is proportional to  $c_1^2$ at long times is consistent with the timescales given above for a single capsid if elongation dominates.

\begin{figure} [hbt]
\epsfig{file=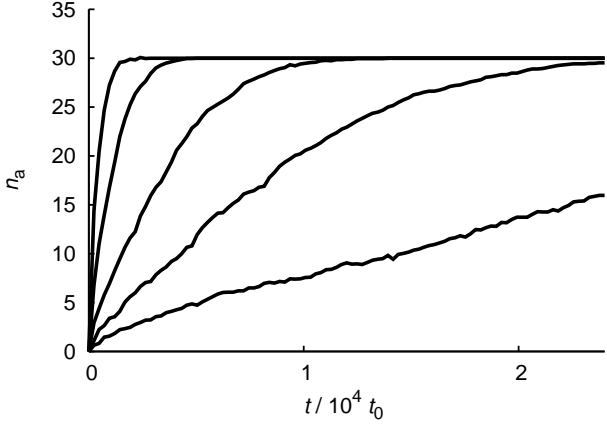,width= \linewidth}
\caption{\label{fig:two}
The time dependence of adsorbed subunits, $\na$, from simulations.  Curves at increasing height correspond to reduced subunit concentrations of $10^3 C_\text{S}= 2.04, 4.07, 8.14, 20.4, 40.7$, with a surface free energy of $\ec = 7$ ($\gc=-5.1$), $\tmax = 1$, and $\eb = 10$.  Simulation results are averaged over 60 independent trajectories.
}
\end{figure}

	\emph{Assembly rates on cores.} As we will see from the simulations described below, the presence of cores modifies the nucleation and elongation timescales and introduces a new one, which describes the adsorption of subunits to the core surface.  In this discussion we assume that cores are commensurate with the size and geometry of capsids; we discuss the general case elsewhere.  If there is no assembly of adsorbed subunits, the adsorption timescale is $\tau_{\text{ad}}= n_{\text{s}}/k_{\text{ad}}c_1$, with $k_\text{ad}$ the adsorption rate constant, and the number of adsorbed subunits at saturation, $n_s$,  can be calculated for Langmuir adsorption
\begin{equation}
\label{eq:csurf}
n_{\text{s}}=V_{\text{C}} c_1 \exp(-\beta g_{\text{c}})/\left(1+c_1 \exp(-\beta g_{\text{c}})\right)
\end{equation}
where $V_{\text{C}}\approx N \sigma^3$ is the volume available to adsorbed subunits, $\beta =1/\kt$ is the inverse of the thermal energy, and $g_c$  is the surface-subunit free energy.  Adsorption will usually be fast compared to assembly rates, which are slow compared to the diffusion limited rate for protein collisions \cite{Zlotnick1999}.  Simulation results demonstrate fast adsorption; Fig.~\ref{fig:two} shows the number of adsorbed subunits, $\na$, as a function of time for several subunit concentrations.  In each case, there is a rapid initial rise in the number of adsorbed subunits, which analysis of trajectories confirms is due to non-specific subunit adsorption (i.e. without binding to other subunits).

For fast adsorption, nucleation takes place at an effective surface concentration of $c_\text{surf}=n_\text{s} \sigma^3 /V_\text{C}$ with a timescale
\begin{equation}
\tau_{\text{nuc}}^{\text{core}}=f_{\text{surf}} c_{\text{surf}}^{-n_{\text{nuc}}}
\label{eq:tauNuc}
\end{equation}
where we define the surface assembly rate constant $f_{\text{surf}}=f D_{\text{C}}/D $ , with $D$ and $D_\text{C}$ the diffusion constants for free and adsorbed subunits, respectively, and we require that assembly rates are proportional to the frequency of subunit-subunit collisions \cite{Berg1977}.  In simulations for this work $D_C=D$ because subunit friction is isotropic, but we will explore the effects of impeded surface diffusion elsewhere. Desorption of a nucleated intermediate is unlikely, since it would require breaking multiple subunit-subunit or subunit-core interactions.  Assembly therefore leads to a positive flux of adsorbed subunits and elongation occurs at roughly the same concentration $c_\text{surf}$ with a timescale
\begin{equation}
\tau_{\text{elong}}^{\text{core}}\approx (N-n_{\text {s}}) /k_{\text{ad}} c_1+ (N-n_{\text{nuc}})/f_{\text{surf}}  c_{\text{surf}}
\label{eq:tauElong}
\end{equation}
The first term on the right-hand side of equation (\ref{eq:tauElong}) is the time for the remaining subunits to adsorb to the core surface while the second term accounts for the elongation reaction time.

\begin{figure} [bt]
\epsfig{file=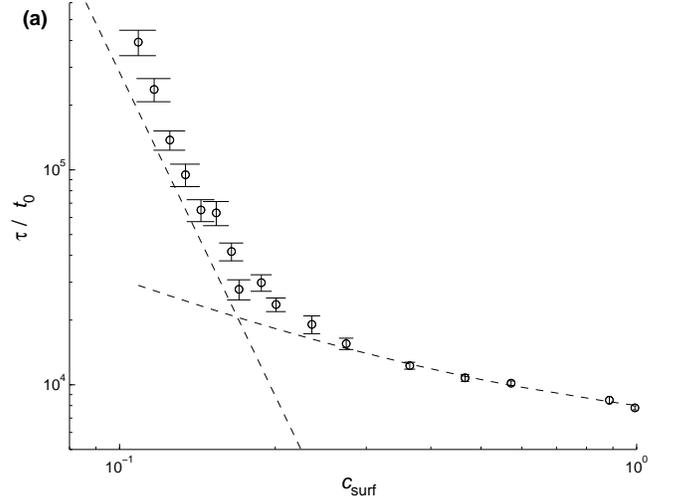,width= \linewidth}
\epsfig{file=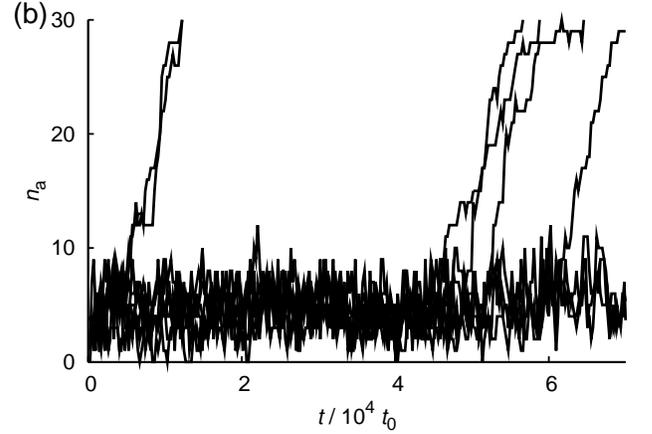,width= \linewidth}

\caption{\label{fig:three}
(a) Average assembly time, $\tau$, for simulations of core encapsidation for varying surface energies plotted as a function of the surface concentration, csurf, described in the text.  Simulated surface energies spanned $\varepsilon_\text{c} \in [4.1,12]$, giving surface free energies of $-\gc \in [2.7,9.6]$.  The upper dashed line is a guide to the eye that indicates a nucleation dominated scaling of $c_\text{surf}^4$ while the lower dashed line is a fit to the rightmost four data points with the form $\tau=A+B/c_\text{surf}$ to illustrate the elongation dominated scaling.  Data points represent an average of 30 or more independent encapsidation trajectories, run at parameter values of $C_\text{S}=8\times 10^{-3}$, $\eb=10$, and $\tmax=1$.  {\bf (b)} Nine individual trajectories are shown for a surface energy in the nucleation regime, $\ec=4.5$ ($c_\text{surf}=0.15$).
}
\end{figure}

Eqs. \ref{eq:tauNuc} and \ref{eq:tauElong} predict that cores enhance assembly rates by a factor
\begin{equation}
\tau/\tau^{\text{core}} = \frac{\tau_{\text{nuc}} + \tau_{\text{elong}}}{N/k_{\text{ad}}c_1 + \tau_{\text{nuc}}^{\text{core}}+\tau_{\text{elong}}^{\text{core}}}
\label{eq:ratio},
\end{equation}
and that the relative timescales for nucleation and elongation can be manipulated by varying the surface-subunit free energy, to yield regimes in which either nucleation or elongation is rate limiting.  Fig.~\ref{fig:three} shows average assembly times for core encapsidation in simulations at varying surface energies $4.1 \le \varepsilon_{\text{c}}/k_{\text{B}}T \le 12 $; the upper and lower dashed lines identify the scaling relations predicted by Eqs. (\ref{eq:tauNuc}) and (\ref{eq:tauElong}) for nucleation and elongation dominated regimes, respectively, and we take $n_\text{nuc} = 5$, although more data would be required to precisely estimate the nucleation size.  In addition, the scaling relation is limited in range because the nucleation size can increase at very low $\ec$.  The nucleus usually corresponds to a small polygon in our simulations (see Fig.~\ref{fig:five}), but the size and geometry of nuclei depend on system parameters.  Time series of $\na$ (the total number of adsorbed subunits) from individual trajectories in the nucleation dominated regime are shown in Fig.~\ref{fig:three}b to illustrate the stochastic nature of the nucleation event.

\begin{figure} [bt]
\epsfig{file=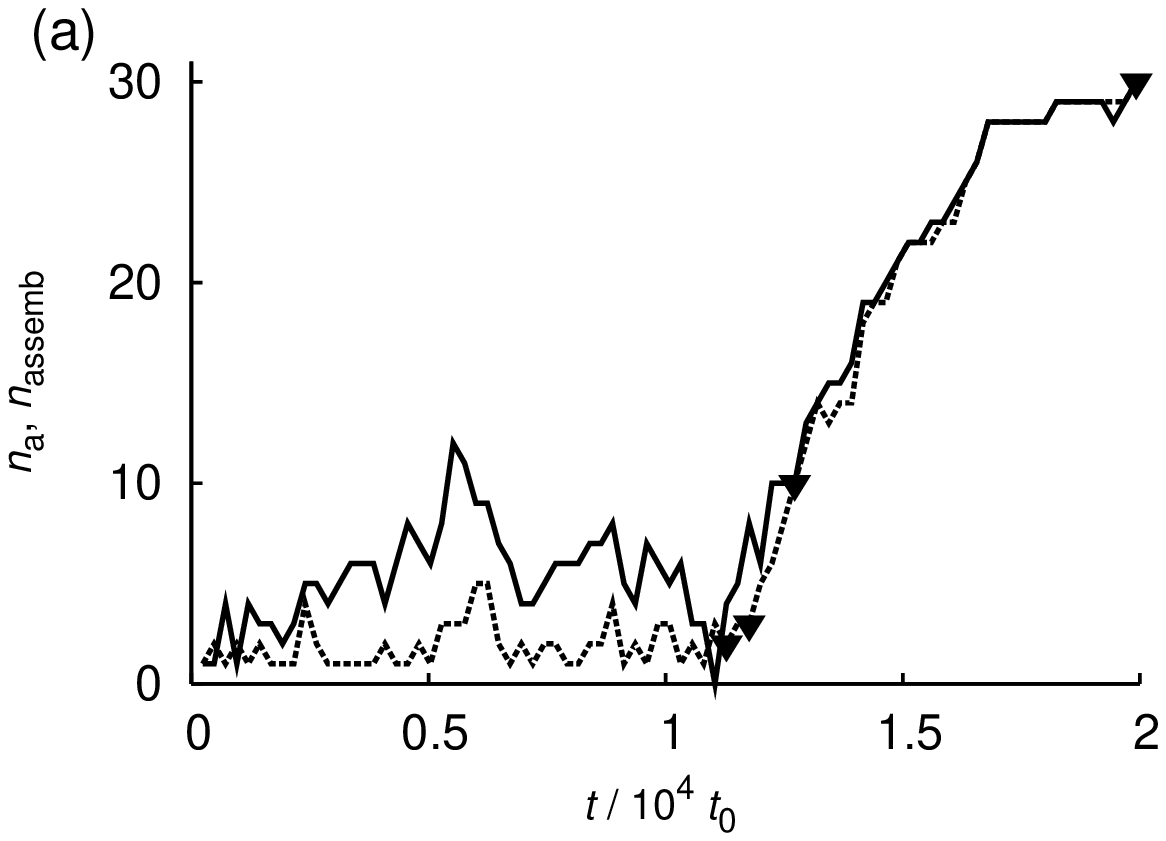,width= \linewidth}
\epsfig{file=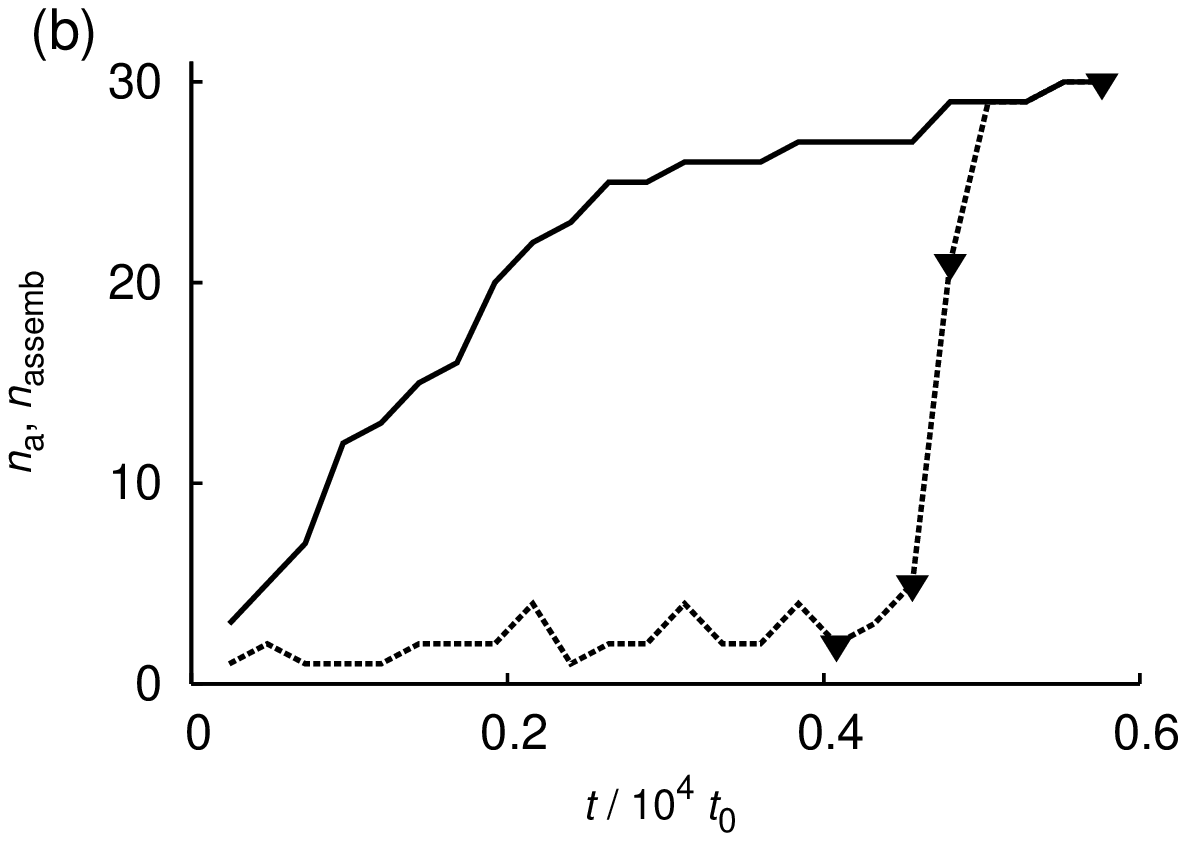,width= \linewidth}
\caption{\label{fig:four}
The time dependences of the total number of adsorbed subunits, $\na$ (solid lines), and the largest assembled cluster, $n_\text{assemb}$ (dashed lines), reveal different assembly mechanisms.  The subunit concentration is $C_\text{S}=8\times 10^{-3}$, and the energy parameters are:  {\bf (a)} low surface energy and moderate binding energy, $\ec=4.5$, $\eb=10$; {\bf (b)} high surface energy and low binding energy, $\ec =12$, $\eb =7$.  The points labeled with ($\blacktriangledown$) in {\bf (a)} and {\bf (b)} correspond to the structures shown in Figs. 5a and 5b, respectively.
}
\end{figure}

\begin{figure} [hbt]
\epsfig{file=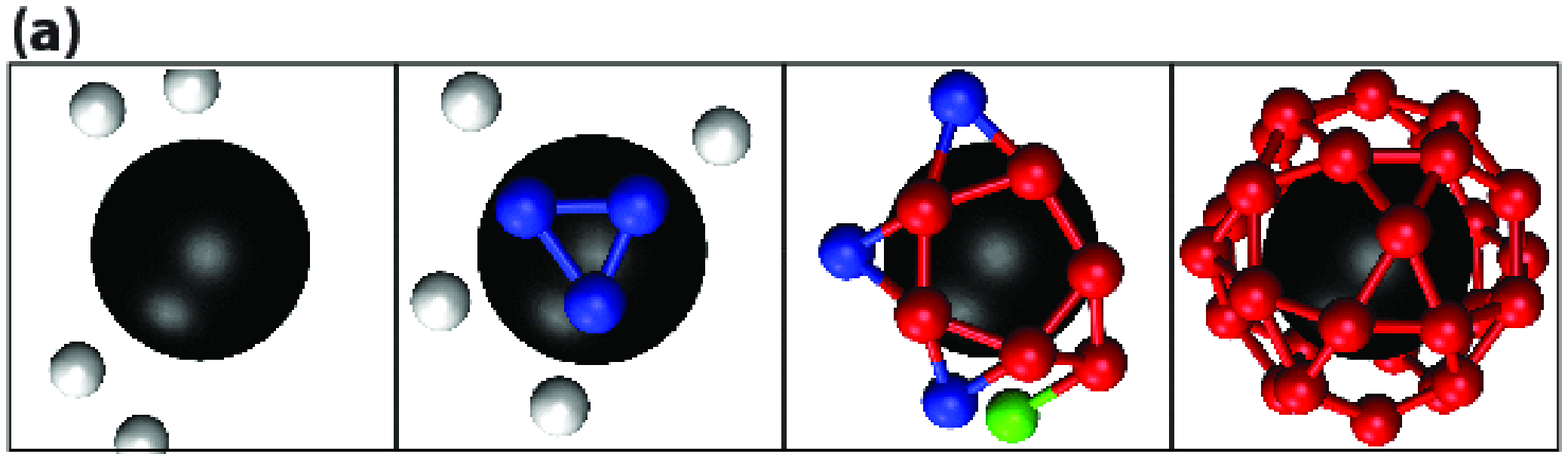,width= \linewidth}
\epsfig{file=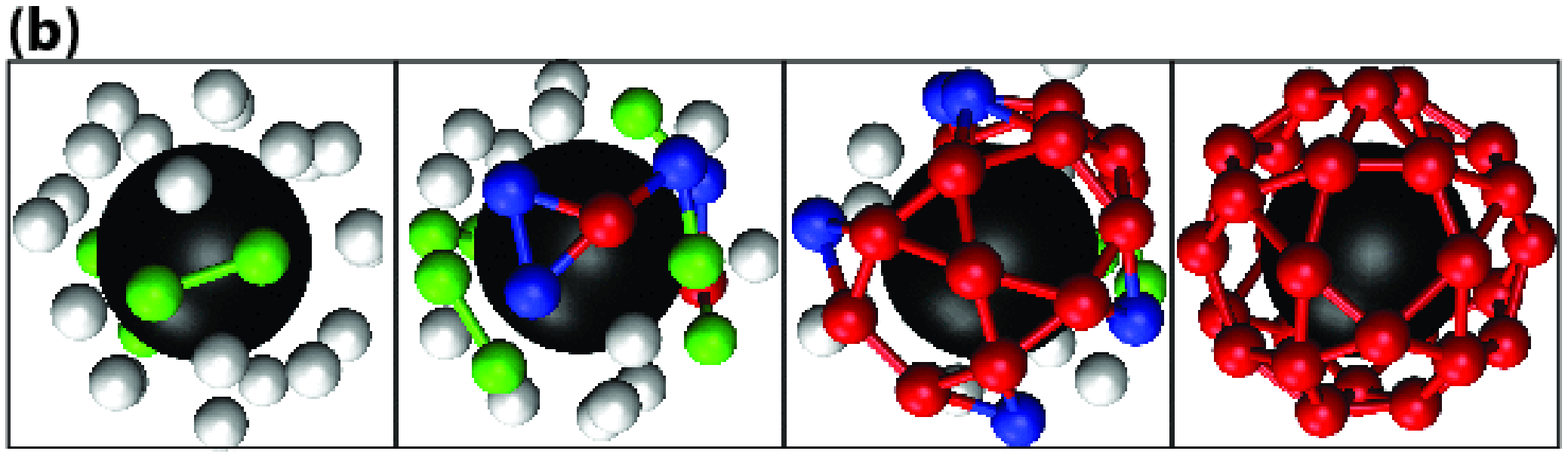,width= \linewidth}
\caption{\label{fig:five}
Snapshots from the simulation trajectories shown in Fig.~\ref{fig:four}a illustrate two assembly mechanisms: {\bf (a)} sequential assembly at low surface energy, $\ec=4.5$, and {\bf (b)} cooperative assembly at high surface enery, $\ec=12$.  Snapshots from right to left correspond to increasing time and correspond to the triangles shown in {\bf (a)} Fig.~\ref{fig:four}a and {\bf (b)} Fig.~\ref{fig:four}c.  The size of subunits is reduced to aid visibility, and subunit color indicates the number of complementary interactions: white, 0; green, 1; blue, 2; red, 3 or 4.  All images of simulation structures in this work were generated with VMD \cite{Humphrey1996}.
}
\end{figure}

\emph{Cooperative assembly.} For large surface free energies, $c_{\text{surf}}\approx 1 $ ,  meaning that adsorption does not saturate until enough subunits have adsorbed to form a capsid, $n_{\text{s}}\approx N$.  Assembly in this regime can occur through a collective reorientation of adsorbed subunits; at low subunit-subunit binding energies ($\eb$) this process typically takes place well after adsorption has saturated and thus resembles the reverse micelle assembly mechanism suggested by McPherson \cite{McPherson2005}.  The time dependences of the number of adsorbed subunits, $\na$, and the number of assembled subunits, $n_\text{assemb}$, are shown for representative trajectories that illustrate the different assembly pathways in Fig.~\ref{fig:four}, and structures from these trajectories are shown in Fig.~\ref{fig:five}.

\begin{figure} [hbt]
\epsfig{file=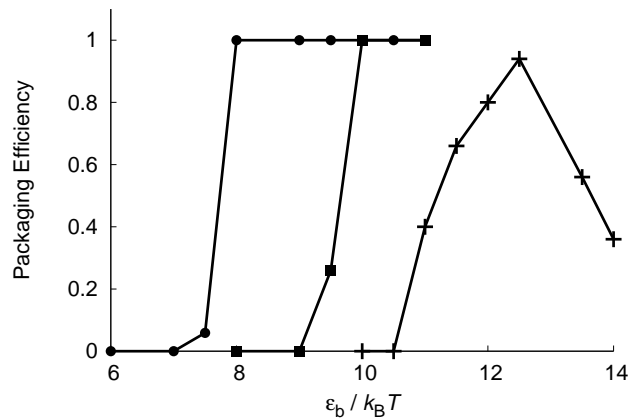,width= \linewidth}
\caption{\label{fig:six}
The efficiency of encapsidation of a model nanoparticle is compared to the fraction of subunits in capsids due to spontaneous assembly of empty capsids, shown with symbol (+).  Packaging efficiencies, or the fraction of independent trajectories in which a nanoparticle was encapsidated by a perfect capsid, are shown for neutral subunits at $\ec=7$ ($\blacksquare$) and $\ec=12$ ($\bullet$), with a subunit concentration of $C_\text{S}=8\times 10^{-3}$ and a final observation time of $t_\text{f} = 48,000$.  Spontaneous assembly results are shown at $C_\text{S}=0.11$ and $t_\text{f} = 600,000$.  The specificity parameter $\tmax=0.5$ for all points.
}
\end{figure}

\subsection{Packaging efficiencies}
In addition to enhancing and controlling rates of assembly, cores can increase assembly yields. As a measure of efficiency of assembly, we observe packaging efficiencies, which are defined as the fraction of independent trajectories for which a nanoparticle is encapsidated by a perfect capsid.  A `perfect capsid' is comprised of 30 subunits, each of which has the maximum number of bonds, 4.  The variation of packaging efficiencies with binding energy is shown in Fig.~\ref{fig:six} for an observation time of $t_f=48,000$, beyond which packaging efficiencies increase only slowly.

{\bf Core control of assembly through heterogeneous nucleation.}
At the subunit concentration considered in Fig.~\ref{fig:six}, $C_\text{S} = 8\times 10^{-3}$, spontaneous assembly into properly formed empty capsids is not observed for any of the binding energies considered, while cores are efficiently encapsidated over relatively wide range of $\eb$.  This observation is consistent with experiments, which find that assembly occurs in the presence of nanoparticles below the critical subunit concentration (CSC) at which spontaneous assembly occurs \cite{Dragnea2007}.  Model nanoparticles enhance assembly because favorable core-subunit interactions lead to a high local concentration, $c_\text{surf}$, of `adsorbed' subunits near core surfaces (see Eq. (\ref{eq:csurf})).  Assembly and encapsidation occur when the effective surface concentration exceeds the CSC.  Core encapsidation simulations were not carried out at binding energies of $\eb > 11$  to ensure that there was no assembly in the bath.  At higher binding energies, spontaneous assembly is rapid and depletes the concentration of free subunits,  and thus suppresses nanoparticle encapsidation and decreases packaging efficiencies.  The competition between core-controlled and spontaneous assembly will be explored in a future work.

Assembly on cores is robust in the sense that packaging efficiencies remain near 100\% over wide ranges of the subunit binding energy $\eb$ and the surface attraction energy, $\ec$.  As a comparison, we consider independent simulations of empty capsid assembly (without model nanoparticles) at a subunit concentration of $C_\text{S}=0.11$ for whichspontaneous assembly is relatively productive. As a measure of assembly effectiveness in empty capsid simulations, we define the `packaging efficiency' as the fraction of subunits in complete capsids.  At an observation time of $t_{\text{f}}=6\times 10^5 $ , approximately 12 times longer than the observation time for core encapsidation, efficient assembly occurs over the relatively narrow range of $11.0 \lesssim \epsilon_{\text{b}} \lesssim 13.5 $.  Empty capsid assembly is thwarted by two forms of kinetic traps at higher values of subunit-subunit binding energies.  If new assembly intermediates form too rapidly, the pool of free subunits becomes depleted before most capsids finish assembling \cite{Ceres2002,Endres2002,Hagan2006,Zhang2006,Zlotnick1994, Zlotnick2000,Zlotnick1999}, and malformed structures result when additional subunits bind more rapidly than strained bonds can anneal within a partial capsid\cite{Hagan2006,Nguyen2007,Schwartz1998,Sorger1986}.  Cores suppress the first of these traps by enabling rapid assembly well below the CSC, so that the number of nucleation sites is controlled by the concentration of cores even during rapid assembly.

\begin{figure} [hbt]
\epsfig{file=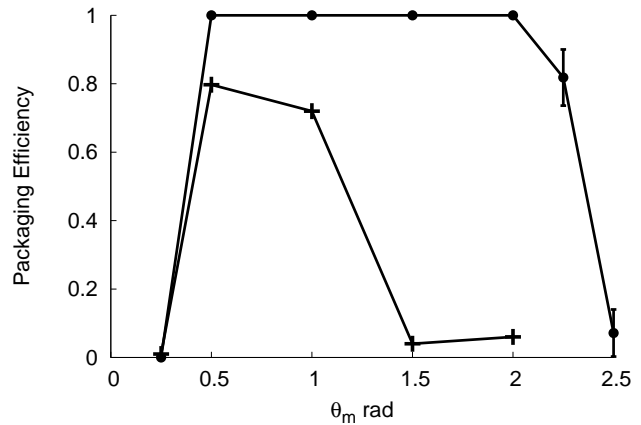,width= \linewidth}
\caption{\label{fig:seven}
{\bf (a)} Variation of assembly with specificity parameter, $\tmax$.  Packaging efficiencies ($\bullet$) are shown for $\eb=10$, $C_\text{S}=8\times 10^{-3}$ and $t_\text{f}=48,000$, while the fractions of subunits in empty capsids (+) are shown for $\eb=12$, $C_\text{S}=0.11$, and. $t_\text{f}=240,000$. 
}
\end{figure}

\begin{figure} [bt]
\epsfig{file=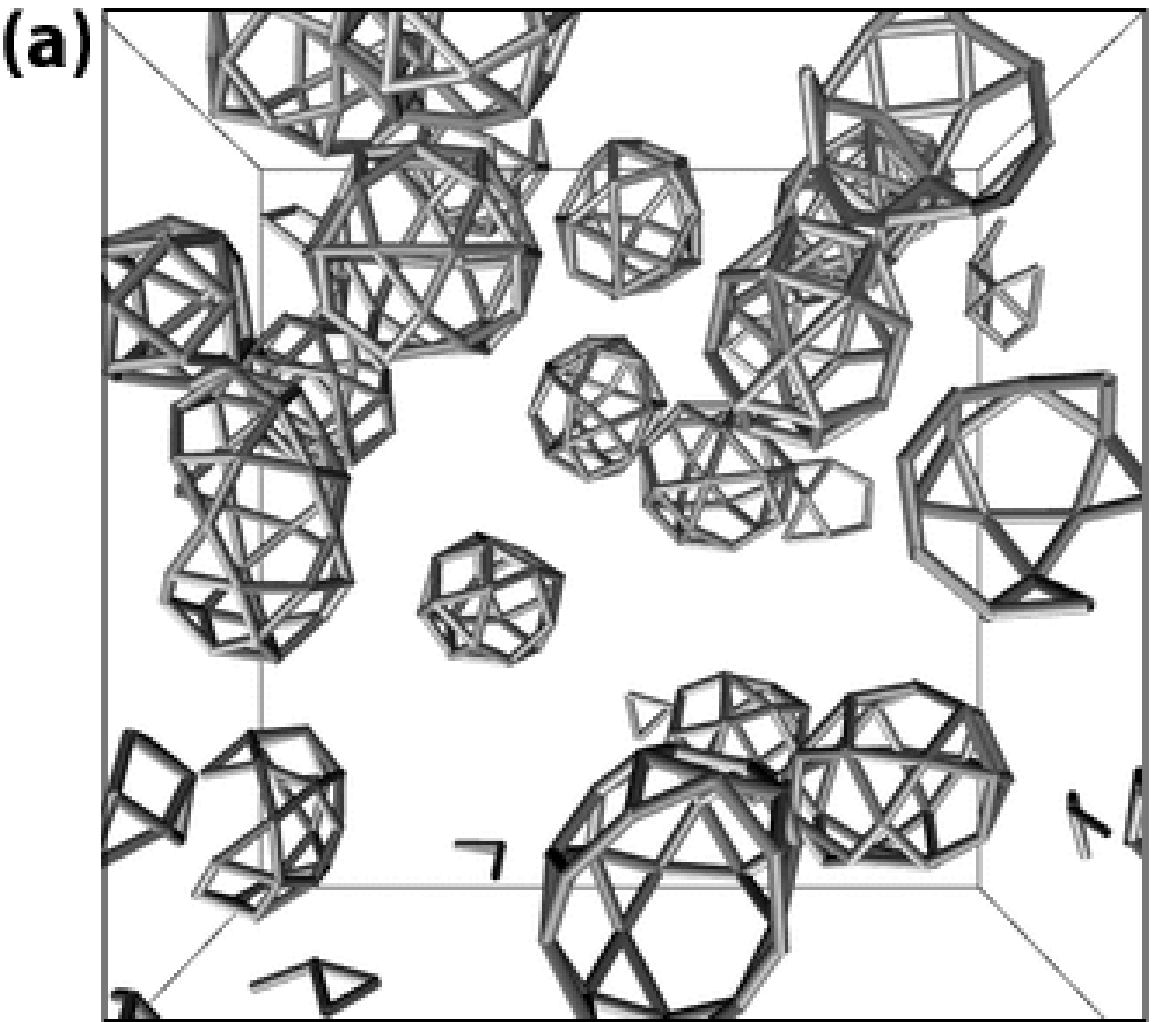,width= .8\linewidth}
\vskip .1in
\epsfig{file=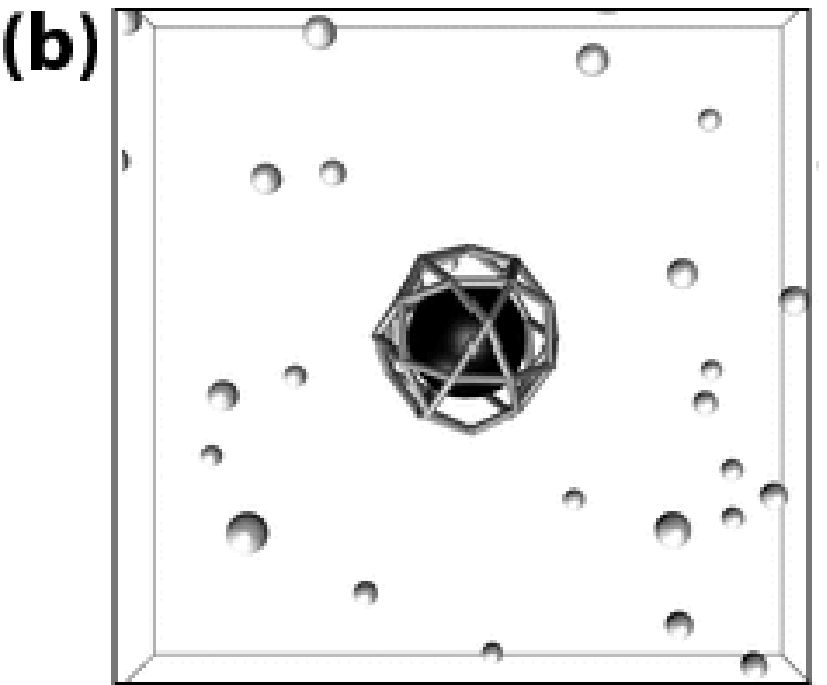,width=.6 \linewidth}
\caption{\label{fig:eight}
 {\bf (a)} Snapshot of part of the simulation box at the end of an empty capsid simulation with $\tmax=2$ for the parameters in Fig.~\ref{fig:seven}.  Some capsids are closed but strained, as indicated by the presence of hexagons or squares.
{\bf (b)} Snapshot at the end of the core encapsidation trajectory shown in Fig.~\ref{fig:five}a.
}
\end{figure}
 
{\bf Core control of assembly through templating.}
The formation of mis-bonded configurations impedes assembly when subunit-subunit binding energies are large compared to $\kt$, because progression from these configurations to a properly formed capsid requires unbinding events and thus is characterized by large activation energies.   An interior core could suppress this form of kinetic trap by acting as a template that directs assembly at all stages towards a morphology consistent with the low free energy capsid.  We explore this capability by varying the subunit specificity parameter $\theta_{\text{m}} $ , which controls the intrinsic likelihood of subunits to form strained bonds (see section \ref{sec:emptyCapsid}).

As shown in Fig.~\ref{fig:seven}, efficient spontaneous assembly of empty capsids occurs over relatively narrow range of $0.5 \lesssim \theta_{\text{m}} \lesssim 1.0 $.  At low values of $\theta_{\text{m}} $ , subunit binding rates are prohibitively slow because most collisions do not lead to bond formation, while at larger values of $\theta_{\text{m}} $ strained bonds tend to be trapped within growing capsids. The presence of assemblages with strained bonds and the lack of free subunits are illustrated by a snapshot from the end of a simulation with $\tmax=2.0$ in Fig.~\ref{fig:eight}a; a snapshot at the end of a core simulation is shown in Fig.~\ref{fig:eight}b for comparison. 

Core controlled assembly, on the other hand, results in packaging efficiencies near 100\% for specificity parameters as large as $\theta_{\text{m}} =2.25$ .  Because the core is commensurate with the size of a perfect capsid, subunits are driven to bind with the correct local curvature at all stages of the assembly process.  Note that the core simulations in Fig.~\ref{fig:seven} benefit from heterogeneous nucleation as well as templating; subunit binding rates and free energies increase with  $\theta_{\text{m}} $, and thus so do spontaneous nucleation rates.  We determine that templating becomes increasingly important as $\tmax$ rises because empty capsid simulations with $\theta_{\text{m}} \geq 1.5$ yield a significant fraction of malformed structures.  Simulations to be presented at a future work, in which there is heterogeneous nucleation but no templating, show a similar degree of sensitivity to $\theta_{\text{m}}$.  

\section{Discussion}
\label{sec:discussion}
In this work we present scaling arguments and simulations that describe the assembly of capsid protein subunits around rigid cores.  The kinetics and efficiency of assembly are predicted as functions of subunit-subunit binding energies, subunit concentrations and subunit-core interaction strengths.  We find that assembly pathways depend sensitively on the strength of subunit-core interactions.  For weak attractions, capsid formation requires assembly of adsorbed subunits into a stable intermediate, followed by sequential adsorption and assembly of individual subunits.  Strong interactions induce rapid adsorption resulting in nearly complete, but disordered, monolayers of subunits, followed by cooperative subunit reordering to form capsids.  Variations in assembly mechanisms are revealed by the time dependence of adsorbed subunits, as shown in Fig.~\ref{fig:two}, and by the scaling of assembly times with subunit concentration, as shown in Fig.~\ref{fig:three}.
	
	Model predictions for the variation of assembly timescales and packaging efficiencies with subunit concentration, $C_\text{S}$, and core-subunit interaction strengths, $\gc$, can be verified in virus-like particle assembly experiments.  The parameter $\gc$ can be related to the experimentally controlled functionalized surface charge density.  Protein adsorption and assembly kinetics can be monitored with time resolved light scattering \cite{Dragnea2007}, and packaging efficiencies can be determined from TEM micrographs \cite{Sun2007}. Successful validation of model predictions would provide strong evidence for the control of assembly mechanisms by a templating component, which prior works suggest is important for understanding viral assembly \emph{in vivo} \cite{Johnson2004,McPherson2005}.  

\emph{Implications for designing and understanding assembly reactions.}  Experiments \cite{Zlotnick2000,Sorger1986} and models \cite{Hagan2006,Endres2002, Nguyen2007, Zlotnick1994,Zhang2006, Wilber2007} show that subunits can spontaneously assemble into low free energy ordered states with high yield and selectivity, but that effective assembly is limited to optimal ranges of the forces that drive assembly.  For non-optimal interactions, assembly is either not thermodynamically favored, or thwarted by long-lived disordered states.  These limitations of spontaneous assembly have shaped the evolution of assembling components in biological systems, and similarly constrain the development of assembled nanostructured materials. Our results suggest that cooperative interactions between disparate assembling components offer the potential to circumvent some limitations of spontaneous assembly, particularly through heterogeneous nucleation and templating.

The novel mechanisms and capabilities of multicomponent assembly introduce new considerations for the design of assembly processes.  For example, Zlotnick and coworkers\cite{Zlotnick1999} show that a slow nucleation step in the spontaneous assembly of empty capsids can suppress kinetic traps.  This condition is met for some parameter values our simulation model -- initial assembly steps can be slow in comparison to later ones because subunits in small intermediates have few bonds.  As subunit-subunit binding energies are decreased these initial steps become even slower; however, the capsid products are less thermodynamically stable and subsequent ``elongation'' rates can also decrease, resulting in increased assembly times.  Introducing cores to the simulations provides an independent means to control the formation of assembly nuclei while still enabling fast elongation kinetics, and thereby promotes rapid assembly with high yield and selectivity.

Controling assembly kinetics by changing the properties of a template component will be useful for designing synthetic or biomimetic assembly reactions for which it is impractical to change the molecular structure of subunits or environmental conditions.  For example, protein-protein interactions, and hence the critical subunit concentration (CSC), can be controlled in virus-like particle experiments by varying the salt concentration or pH \cite{Zlotnick2000,Ceres2002,Kegel2004}.  Capsid proteins denature, however, if these parameters are changed too far from physiological conditions.  Our results suggest that varying the functionalized surface charge density on nanoparticles enables independent control over the CSC and packaging efficiencies. Similarly, viruses have limited capability to control the cellular environments in which they replicate, and amino acids at capsid protein-protein interfaces are highly conserved, perhaps in part because the need for capsid dissociation upon infecting a new cell constrains these interactions.    Interactions between capsid proteins and the viral genome or host cellular compenents may provide important alternative avenues to promote and control assembly, and hence viral replication.

\section{Outlook}
\label{sec:outlook}
\emph{Core-assembly geometry incompatibility.}  Our models describe encapsidation of cores with shapes and sizes commensurate with the low free energy capsid product.  The simulation results demonstrate that the influence of core curvature on local subunit-subunit bonding configurations can dramatically influence global capsid morphologies (see Fig.~\ref{fig:seven}).  Core curvature that is inconsistent with the lowest free energy subunit-subunit bonding configurations will introduce frustration and thus may limit the robustness of assembly.  Experimental observations that solid cores \cite{Sun2007} and nucleic acid cores \cite{Krol1999} with different sizes promote assembly of different capsid morphologies demonstrate that frustration is an important consideration in biological and nanostructured assembly processes.  Comparison of our current model with experimental results will identify frustration and future work will explicitly address geometrical frustration.  Additionally, physio-adsorption of subunits on core surfaces could impede lateral diffusion and thereby promote kinetic traps.  However, simulations in which adsorbed subunits had friction constants increased by a factor of 100 in directions tangential to the core surface demonstrated only a small increase in propensity for kinetic traps, although net assembly rates were slower.

\emph{Fluctuating cores.}  Cores comprised of nucleic acids or other macromolecules with dynamic configurations can change size and shape during encapsidation.  Although the timescale arguments presented in Eqs.~\ref{eq:tauNuc} and \ref{eq:tauElong} can be generalized to include an additional timescale that represents core dynamics, the possibility of additional forms of kinetic traps due to fluctuating core configurations should be explored. 

\section{Conclusions}
In summary, we have presented simulations and scaling arguments that describe the assembly of solubilized subunits around rigid cores.  These models mimic the dynamical assembly of viral capsid proteins around functionalized inorganic nanoparticles, but are general enough for broad applicability in describing the assembly of biological or nanostructured materials around templates.  We find that template properties can dramatically influence assembly timescales and mechanisms, as evidenced by the prediction of a novel assembly mechanism not seen during the assembly of empty capsids.  These conclusions may be significant for understanding the role of nucleic acids in the viral assembly and for designing nanomaterials or drug delivery vehicles to interact with cargo molecules.  

{\bf Acknowledgments} I gratefully acknowledge Rob Jack and Bogdan Dragnea for insightful discussions during this work, Aaron Dinner, Bulbul Chakraborty, and Jane' Kondev for critical readings of the manuscript, and Oren Elrad for assistance in preparing the manuscript.  Funding was provided by an HHMI-NIBIB Interfaces Initiative grant to Brandeis University and Brandeis University startup funds.

\begin{thebibliography}{79}
\expandafter\ifx\csname natexlab\endcsname\relax\def\natexlab#1{#1}\fi
\expandafter\ifx\csname bibnamefont\endcsname\relax
  \def\bibnamefont#1{#1}\fi
\expandafter\ifx\csname bibfnamefont\endcsname\relax
  \def\bibfnamefont#1{#1}\fi
\expandafter\ifx\csname citenamefont\endcsname\relax
  \def\citenamefont#1{#1}\fi
\expandafter\ifx\csname url\endcsname\relax
  \def\url#1{\texttt{#1}}\fi
\expandafter\ifx\csname urlprefix\endcsname\relax\def\urlprefix{URL }\fi
\providecommand{\bibinfo}[2]{#2}
\providecommand{\eprint}[2][]{\url{#2}}

\bibitem[{\citenamefont{Glotzer}(2004)}]{Glotzer2004}
\bibinfo{author}{\bibfnamefont{S.~C.} \bibnamefont{Glotzer}},
  \bibinfo{journal}{Science} \textbf{\bibinfo{volume}{306}},
  \bibinfo{pages}{419} (\bibinfo{year}{2004}).

\bibitem[{\citenamefont{Whitesides and Grzybowski}(2002)}]{Whitesides2002b}
\bibinfo{author}{\bibfnamefont{G.~M.} \bibnamefont{Whitesides}}
  \bibnamefont{and}
  \bibinfo{author}{\bibfnamefont{B.}~\bibnamefont{Grzybowski}},
  \bibinfo{journal}{Science} \textbf{\bibinfo{volume}{295}},
  \bibinfo{pages}{2418} (\bibinfo{year}{2002}).

\bibitem[{\citenamefont{Douglas and Young}(2006)}]{Douglas2006}
\bibinfo{author}{\bibfnamefont{T.}~\bibnamefont{Douglas}} \bibnamefont{and}
  \bibinfo{author}{\bibfnamefont{M.}~\bibnamefont{Young}},
  \bibinfo{journal}{Science} \textbf{\bibinfo{volume}{312}},
  \bibinfo{pages}{873(3)} (\bibinfo{year}{2006}).

\bibitem[{\citenamefont{Valery et~al.}(2003)\citenamefont{Valery, Paternostre,
  Robert, Gulik-Krzywicki, Narayanan, Dedieu, Keller, Torres, Cherif-Cheikh,
  Calvo et~al.}}]{Valery2003}
\bibinfo{author}{\bibfnamefont{C.}~\bibnamefont{Valery}},
  \bibinfo{author}{\bibfnamefont{M.}~\bibnamefont{Paternostre}},
  \bibinfo{author}{\bibfnamefont{B.}~\bibnamefont{Robert}},
  \bibinfo{author}{\bibfnamefont{T.}~\bibnamefont{Gulik-Krzywicki}},
  \bibinfo{author}{\bibfnamefont{T.}~\bibnamefont{Narayanan}},
  \bibinfo{author}{\bibfnamefont{J.-C.} \bibnamefont{Dedieu}},
  \bibinfo{author}{\bibfnamefont{G.}~\bibnamefont{Keller}},
  \bibinfo{author}{\bibfnamefont{M.-L.} \bibnamefont{Torres}},
  \bibinfo{author}{\bibfnamefont{R.}~\bibnamefont{Cherif-Cheikh}},
  \bibinfo{author}{\bibfnamefont{P.}~\bibnamefont{Calvo}},
  \bibnamefont{et~al.}, \bibinfo{journal}{Proc. Natl. Acad. Sci. U. S. A.}
  \textbf{\bibinfo{volume}{100}}, \bibinfo{pages}{10258}
  (\bibinfo{year}{2003}).

\bibitem[{\citenamefont{Yan et~al.}(2003)\citenamefont{Yan, Park, Finkelstein,
  Reif, and LaBean}}]{Yan2003}
\bibinfo{author}{\bibfnamefont{H.}~\bibnamefont{Yan}},
  \bibinfo{author}{\bibfnamefont{S.~H.} \bibnamefont{Park}},
  \bibinfo{author}{\bibfnamefont{G.}~\bibnamefont{Finkelstein}},
  \bibinfo{author}{\bibfnamefont{J.~H.} \bibnamefont{Reif}}, \bibnamefont{and}
  \bibinfo{author}{\bibfnamefont{T.~H.} \bibnamefont{LaBean}},
  \bibinfo{journal}{Science} \textbf{\bibinfo{volume}{301}},
  \bibinfo{pages}{1882} (\bibinfo{year}{2003}).

\bibitem[{\citenamefont{Strable et~al.}(2004)\citenamefont{Strable, Johnson,
  and Finn}}]{Strable2004}
\bibinfo{author}{\bibfnamefont{E.}~\bibnamefont{Strable}},
  \bibinfo{author}{\bibfnamefont{J.~E.} \bibnamefont{Johnson}},
  \bibnamefont{and} \bibinfo{author}{\bibfnamefont{M.~G.} \bibnamefont{Finn}},
  \bibinfo{journal}{Nano Letters} \textbf{\bibinfo{volume}{4}},
  \bibinfo{pages}{1385} (\bibinfo{year}{2004}).

\bibitem[{\citenamefont{Blum et~al.}(2004)\citenamefont{Blum, Soto, Wilson,
  Cole, Kim, Gnade, Chatterji, Ochoa, Lin, Johnson et~al.}}]{Blum2004}
\bibinfo{author}{\bibfnamefont{A.~S.} \bibnamefont{Blum}},
  \bibinfo{author}{\bibfnamefont{C.~M.} \bibnamefont{Soto}},
  \bibinfo{author}{\bibfnamefont{C.~D.} \bibnamefont{Wilson}},
  \bibinfo{author}{\bibfnamefont{J.~D.} \bibnamefont{Cole}},
  \bibinfo{author}{\bibfnamefont{M.}~\bibnamefont{Kim}},
  \bibinfo{author}{\bibfnamefont{B.}~\bibnamefont{Gnade}},
  \bibinfo{author}{\bibfnamefont{A.}~\bibnamefont{Chatterji}},
  \bibinfo{author}{\bibfnamefont{W.~F.} \bibnamefont{Ochoa}},
  \bibinfo{author}{\bibfnamefont{T.~W.} \bibnamefont{Lin}},
  \bibinfo{author}{\bibfnamefont{J.~E.} \bibnamefont{Johnson}},
  \bibnamefont{et~al.}, \bibinfo{journal}{Nano Letters}
  \textbf{\bibinfo{volume}{4}}, \bibinfo{pages}{867} (\bibinfo{year}{2004}).

\bibitem[{\citenamefont{Johnson et~al.}(2005)\citenamefont{Johnson, Tang,
  Nyame, Willits, Young, and Zlotnick}}]{Johnson2005}
\bibinfo{author}{\bibfnamefont{J.~M.} \bibnamefont{Johnson}},
  \bibinfo{author}{\bibfnamefont{J.~H.} \bibnamefont{Tang}},
  \bibinfo{author}{\bibfnamefont{Y.}~\bibnamefont{Nyame}},
  \bibinfo{author}{\bibfnamefont{D.}~\bibnamefont{Willits}},
  \bibinfo{author}{\bibfnamefont{M.~J.} \bibnamefont{Young}}, \bibnamefont{and}
  \bibinfo{author}{\bibfnamefont{A.}~\bibnamefont{Zlotnick}},
  \bibinfo{journal}{Nano Letters} \textbf{\bibinfo{volume}{5}},
  \bibinfo{pages}{765} (\bibinfo{year}{2005}).

\bibitem[{\citenamefont{Casini et~al.}(2004)\citenamefont{Casini, Graham,
  Heine, Garcea, and Wu}}]{Casini2004}
\bibinfo{author}{\bibfnamefont{G.~L.} \bibnamefont{Casini}},
  \bibinfo{author}{\bibfnamefont{D.}~\bibnamefont{Graham}},
  \bibinfo{author}{\bibfnamefont{D.}~\bibnamefont{Heine}},
  \bibinfo{author}{\bibfnamefont{R.~L.} \bibnamefont{Garcea}},
  \bibnamefont{and} \bibinfo{author}{\bibfnamefont{D.~T.} \bibnamefont{Wu}},
  \bibinfo{journal}{Virology} \textbf{\bibinfo{volume}{325}},
  \bibinfo{pages}{320} (\bibinfo{year}{2004}).

\bibitem[{\citenamefont{Singh and Zlotnick}(2003)}]{Singh2003}
\bibinfo{author}{\bibfnamefont{S.}~\bibnamefont{Singh}} \bibnamefont{and}
  \bibinfo{author}{\bibfnamefont{A.}~\bibnamefont{Zlotnick}},
  \bibinfo{journal}{J. Biol. Chem.} \textbf{\bibinfo{volume}{278}},
  \bibinfo{pages}{18249} (\bibinfo{year}{2003}).

\bibitem[{\citenamefont{Willits et~al.}(2003)\citenamefont{Willits, Zhao,
  Olson, Baker, Zlotnick, Johnson, Douglas, and Young}}]{Willits2003}
\bibinfo{author}{\bibfnamefont{D.}~\bibnamefont{Willits}},
  \bibinfo{author}{\bibfnamefont{X.}~\bibnamefont{Zhao}},
  \bibinfo{author}{\bibfnamefont{N.}~\bibnamefont{Olson}},
  \bibinfo{author}{\bibfnamefont{T.~S.} \bibnamefont{Baker}},
  \bibinfo{author}{\bibfnamefont{A.}~\bibnamefont{Zlotnick}},
  \bibinfo{author}{\bibfnamefont{J.~E.} \bibnamefont{Johnson}},
  \bibinfo{author}{\bibfnamefont{T.}~\bibnamefont{Douglas}}, \bibnamefont{and}
  \bibinfo{author}{\bibfnamefont{M.~J.} \bibnamefont{Young}},
  \bibinfo{journal}{Virology} \textbf{\bibinfo{volume}{306}},
  \bibinfo{pages}{280} (\bibinfo{year}{2003}).

\bibitem[{\citenamefont{Zlotnick et~al.}(2000)\citenamefont{Zlotnick, Aldrich,
  Johnson, Ceres, and Young}}]{Zlotnick2000}
\bibinfo{author}{\bibfnamefont{A.}~\bibnamefont{Zlotnick}},
  \bibinfo{author}{\bibfnamefont{R.}~\bibnamefont{Aldrich}},
  \bibinfo{author}{\bibfnamefont{J.~M.} \bibnamefont{Johnson}},
  \bibinfo{author}{\bibfnamefont{P.}~\bibnamefont{Ceres}}, \bibnamefont{and}
  \bibinfo{author}{\bibfnamefont{M.~J.} \bibnamefont{Young}},
  \bibinfo{journal}{Virology} \textbf{\bibinfo{volume}{277}},
  \bibinfo{pages}{450} (\bibinfo{year}{2000}).

\bibitem[{\citenamefont{Klug}(1999)}]{Klug1999}
\bibinfo{author}{\bibfnamefont{A.}~\bibnamefont{Klug}},
  \bibinfo{journal}{Philos. Trans. R. Soc. Lond. B. Biol. Sci.}
  \textbf{\bibinfo{volume}{354}}, \bibinfo{pages}{531} (\bibinfo{year}{1999}).

\bibitem[{\citenamefont{Zlotnick et~al.}(1996)\citenamefont{Zlotnick, Cheng,
  Conway, Booy, Steven, Stahl, and Wingfield}}]{Zlotnick1996}
\bibinfo{author}{\bibfnamefont{A.}~\bibnamefont{Zlotnick}},
  \bibinfo{author}{\bibfnamefont{N.}~\bibnamefont{Cheng}},
  \bibinfo{author}{\bibfnamefont{J.~F.} \bibnamefont{Conway}},
  \bibinfo{author}{\bibfnamefont{F.~P.} \bibnamefont{Booy}},
  \bibinfo{author}{\bibfnamefont{A.~C.} \bibnamefont{Steven}},
  \bibinfo{author}{\bibfnamefont{S.~J.} \bibnamefont{Stahl}}, \bibnamefont{and}
  \bibinfo{author}{\bibfnamefont{P.~T.} \bibnamefont{Wingfield}},
  \bibinfo{journal}{Biochemistry} \textbf{\bibinfo{volume}{35}},
  \bibinfo{pages}{7412} (\bibinfo{year}{1996}).

\bibitem[{\citenamefont{Fox et~al.}(1994)\citenamefont{Fox, Johnson, and
  Young}}]{Fox1994}
\bibinfo{author}{\bibfnamefont{J.~M.} \bibnamefont{Fox}},
  \bibinfo{author}{\bibfnamefont{J.~E.} \bibnamefont{Johnson}},
  \bibnamefont{and} \bibinfo{author}{\bibfnamefont{M.~J.} \bibnamefont{Young}},
  \bibinfo{journal}{Seminars in Virology} \textbf{\bibinfo{volume}{5}},
  \bibinfo{pages}{51} (\bibinfo{year}{1994}).

\bibitem[{\citenamefont{Butler and Klug}(1978)}]{Butler1978}
\bibinfo{author}{\bibfnamefont{P.~J.~G.} \bibnamefont{Butler}}
  \bibnamefont{and} \bibinfo{author}{\bibfnamefont{A.}~\bibnamefont{Klug}},
  \bibinfo{journal}{Sci. Am.} \textbf{\bibinfo{volume}{239}},
  \bibinfo{pages}{62} (\bibinfo{year}{1978}).

\bibitem[{\citenamefont{Fraenkelconrat and
  Williams}(1955)}]{Fraenkelconrat1955}
\bibinfo{author}{\bibfnamefont{H.}~\bibnamefont{Fraenkelconrat}}
  \bibnamefont{and} \bibinfo{author}{\bibfnamefont{R.~C.}
  \bibnamefont{Williams}}, \bibinfo{journal}{Proc. Natl. Acad. Sci. U. S. A.}
  \textbf{\bibinfo{volume}{41}}, \bibinfo{pages}{690} (\bibinfo{year}{1955}).

\bibitem[{\citenamefont{Crick and Watson}(1956)}]{Crick1956}
\bibinfo{author}{\bibfnamefont{F.~H.~C.} \bibnamefont{Crick}} \bibnamefont{and}
  \bibinfo{author}{\bibfnamefont{J.~D.} \bibnamefont{Watson}},
  \bibinfo{journal}{Nature} \textbf{\bibinfo{volume}{177}},
  \bibinfo{pages}{473} (\bibinfo{year}{1956}).

\bibitem[{\citenamefont{Caspar and Klug}(1962)}]{Caspar1962}
\bibinfo{author}{\bibfnamefont{D.~L.~D.} \bibnamefont{Caspar}}
  \bibnamefont{and} \bibinfo{author}{\bibfnamefont{A.}~\bibnamefont{Klug}},
  \bibinfo{journal}{Cold Spring Harbor Symp. Quant. Biol.}
  \textbf{\bibinfo{volume}{27}}, \bibinfo{pages}{1} (\bibinfo{year}{1962}).

\bibitem[{\citenamefont{Berger et~al.}(1994)\citenamefont{Berger, Shor,
  Tuckerkellogg, and King}}]{Berger1994}
\bibinfo{author}{\bibfnamefont{B.}~\bibnamefont{Berger}},
  \bibinfo{author}{\bibfnamefont{P.~W.} \bibnamefont{Shor}},
  \bibinfo{author}{\bibfnamefont{L.}~\bibnamefont{Tuckerkellogg}},
  \bibnamefont{and} \bibinfo{author}{\bibfnamefont{J.}~\bibnamefont{King}},
  \bibinfo{journal}{Proc. Natl. Acad. Sci. U. S. A.}
  \textbf{\bibinfo{volume}{91}}, \bibinfo{pages}{7732} (\bibinfo{year}{1994}).

\bibitem[{\citenamefont{Zlotnick}(1994)}]{Zlotnick1994}
\bibinfo{author}{\bibfnamefont{A.}~\bibnamefont{Zlotnick}},
  \bibinfo{journal}{J. Mol. Biol.} \textbf{\bibinfo{volume}{241}},
  \bibinfo{pages}{59} (\bibinfo{year}{1994}).

\bibitem[{\citenamefont{Bruinsma et~al.}(2003)\citenamefont{Bruinsma, Gelbart,
  Reguera, Rudnick, and Zandi}}]{Bruinsma2003}
\bibinfo{author}{\bibfnamefont{R.~F.} \bibnamefont{Bruinsma}},
  \bibinfo{author}{\bibfnamefont{W.~M.} \bibnamefont{Gelbart}},
  \bibinfo{author}{\bibfnamefont{D.}~\bibnamefont{Reguera}},
  \bibinfo{author}{\bibfnamefont{J.}~\bibnamefont{Rudnick}}, \bibnamefont{and}
  \bibinfo{author}{\bibfnamefont{R.}~\bibnamefont{Zandi}},
  \bibinfo{journal}{Phys. Rev. Lett.} \textbf{\bibinfo{volume}{90}},
  \bibinfo{pages}{248101} (\bibinfo{year}{2003}).

\bibitem[{\citenamefont{Zandi et~al.}(2004)\citenamefont{Zandi, Reguera,
  Bruinsma, Gelbart, and Rudnick}}]{Zandi2004}
\bibinfo{author}{\bibfnamefont{R.}~\bibnamefont{Zandi}},
  \bibinfo{author}{\bibfnamefont{D.}~\bibnamefont{Reguera}},
  \bibinfo{author}{\bibfnamefont{R.~F.} \bibnamefont{Bruinsma}},
  \bibinfo{author}{\bibfnamefont{W.~M.} \bibnamefont{Gelbart}},
  \bibnamefont{and} \bibinfo{author}{\bibfnamefont{J.}~\bibnamefont{Rudnick}},
  \bibinfo{journal}{Proc. Natl. Acad. Sci. U. S. A.}
  \textbf{\bibinfo{volume}{101}}, \bibinfo{pages}{15556}
  (\bibinfo{year}{2004}).

\bibitem[{\citenamefont{Keef et~al.}(2005)\citenamefont{Keef, Taormina, and
  Twarock}}]{Keef2005}
\bibinfo{author}{\bibfnamefont{T.}~\bibnamefont{Keef}},
  \bibinfo{author}{\bibfnamefont{A.}~\bibnamefont{Taormina}}, \bibnamefont{and}
  \bibinfo{author}{\bibfnamefont{R.}~\bibnamefont{Twarock}},
  \bibinfo{journal}{Physical Biology} \textbf{\bibinfo{volume}{2}},
  \bibinfo{pages}{175} (\bibinfo{year}{2005}).

\bibitem[{\citenamefont{Twarock}(2005)}]{Twarock2005}
\bibinfo{author}{\bibfnamefont{R.}~\bibnamefont{Twarock}},
  \bibinfo{journal}{Bull. Math. Biol.} \textbf{\bibinfo{volume}{67}},
  \bibinfo{pages}{973} (\bibinfo{year}{2005}).

\bibitem[{\citenamefont{Hagan and Chandler}(2006)}]{Hagan2006}
\bibinfo{author}{\bibfnamefont{M.~F.} \bibnamefont{Hagan}} \bibnamefont{and}
  \bibinfo{author}{\bibfnamefont{D.}~\bibnamefont{Chandler}},
  \bibinfo{journal}{Biophys. J.} \textbf{\bibinfo{volume}{91}},
  \bibinfo{pages}{42} (\bibinfo{year}{2006}).

\bibitem[{\citenamefont{Chen and Glotzer}(2007)}]{Chen2007}
\bibinfo{author}{\bibfnamefont{T.}~\bibnamefont{Chen}} \bibnamefont{and}
  \bibinfo{author}{\bibfnamefont{S.~C.} \bibnamefont{Glotzer}},
  \bibinfo{journal}{Physical Review E.} \textbf{\bibinfo{volume}{75}},
  \bibinfo{pages}{051504} (\bibinfo{year}{2007}).

\bibitem[{\citenamefont{Nguyen et~al.}(2007)\citenamefont{Nguyen, Reddy, and
  Brooks}}]{Nguyen2007}
\bibinfo{author}{\bibfnamefont{H.~D.} \bibnamefont{Nguyen}},
  \bibinfo{author}{\bibfnamefont{V.~S.} \bibnamefont{Reddy}}, \bibnamefont{and}
  \bibinfo{author}{\bibfnamefont{C.~L.} \bibnamefont{Brooks}},
  \bibinfo{journal}{Nano Letters} \textbf{\bibinfo{volume}{7}},
  \bibinfo{pages}{338} (\bibinfo{year}{2007}).

\bibitem[{\citenamefont{Kegel and van~der Schoot}(2004)}]{Kegel2004}
\bibinfo{author}{\bibfnamefont{W.~K.} \bibnamefont{Kegel}} \bibnamefont{and}
  \bibinfo{author}{\bibfnamefont{P.}~\bibnamefont{van~der Schoot}},
  \bibinfo{journal}{Biophys. J.} \textbf{\bibinfo{volume}{86}},
  \bibinfo{pages}{3905} (\bibinfo{year}{2004}).

\bibitem[{\citenamefont{van~der Schoot and Zandi}(2007)}]{vanderSchoot2007}
\bibinfo{author}{\bibfnamefont{P.}~\bibnamefont{van~der Schoot}}
  \bibnamefont{and} \bibinfo{author}{\bibfnamefont{R.}~\bibnamefont{Zandi}},
  \bibinfo{journal}{Physical Biology} p. \bibinfo{pages}{296}
  (\bibinfo{year}{2007}).

\bibitem[{\citenamefont{Hicks and Henley}(2006)}]{Hicks2006}
\bibinfo{author}{\bibfnamefont{S.~D.} \bibnamefont{Hicks}} \bibnamefont{and}
  \bibinfo{author}{\bibfnamefont{C.~L.} \bibnamefont{Henley}},
  \bibinfo{journal}{Phys. Rev. E} \textbf{\bibinfo{volume}{74}},
  \bibinfo{pages}{031912} (\bibinfo{year}{2006}).

\bibitem[{\citenamefont{Dimmock et~al.}(2001)\citenamefont{Dimmock, Easton, and
  Leppard}}]{Dimmock2001}
\bibinfo{author}{\bibfnamefont{N.}~\bibnamefont{Dimmock}},
  \bibinfo{author}{\bibfnamefont{A.}~\bibnamefont{Easton}}, \bibnamefont{and}
  \bibinfo{author}{\bibfnamefont{K.}~\bibnamefont{Leppard}},
  \emph{\bibinfo{title}{Introduction to modern virology}}
  (\bibinfo{publisher}{Blackwell Publishing}, \bibinfo{address}{Malden, MA},
  \bibinfo{year}{2001}).

\bibitem[{\citenamefont{Sun et~al.}(2007)\citenamefont{Sun, DuFort, Daniel,
  Murali, Chen, Gopinath, Stein, De, Rotello, Holzenburg et~al.}}]{Sun2007}
\bibinfo{author}{\bibfnamefont{J.}~\bibnamefont{Sun}},
  \bibinfo{author}{\bibfnamefont{C.}~\bibnamefont{DuFort}},
  \bibinfo{author}{\bibfnamefont{M.~C.} \bibnamefont{Daniel}},
  \bibinfo{author}{\bibfnamefont{A.}~\bibnamefont{Murali}},
  \bibinfo{author}{\bibfnamefont{C.}~\bibnamefont{Chen}},
  \bibinfo{author}{\bibfnamefont{K.}~\bibnamefont{Gopinath}},
  \bibinfo{author}{\bibfnamefont{B.}~\bibnamefont{Stein}},
  \bibinfo{author}{\bibfnamefont{M.}~\bibnamefont{De}},
  \bibinfo{author}{\bibfnamefont{V.~M.} \bibnamefont{Rotello}},
  \bibinfo{author}{\bibfnamefont{A.}~\bibnamefont{Holzenburg}},
  \bibnamefont{et~al.}, \bibinfo{journal}{Proc. Natl. Acad. Sci. U. S. A.}
  \textbf{\bibinfo{volume}{104}}, \bibinfo{pages}{1354} (\bibinfo{year}{2007}).

\bibitem[{\citenamefont{Dixit et~al.}(2006)\citenamefont{Dixit, Goicochea,
  Daniel, Murali, Bronstein, De, Stein, Rotello, Kao, and Dragnea}}]{Dixit2006}
\bibinfo{author}{\bibfnamefont{S.~K.} \bibnamefont{Dixit}},
  \bibinfo{author}{\bibfnamefont{N.~L.} \bibnamefont{Goicochea}},
  \bibinfo{author}{\bibfnamefont{M.~C.} \bibnamefont{Daniel}},
  \bibinfo{author}{\bibfnamefont{A.}~\bibnamefont{Murali}},
  \bibinfo{author}{\bibfnamefont{L.}~\bibnamefont{Bronstein}},
  \bibinfo{author}{\bibfnamefont{M.}~\bibnamefont{De}},
  \bibinfo{author}{\bibfnamefont{B.}~\bibnamefont{Stein}},
  \bibinfo{author}{\bibfnamefont{V.~M.} \bibnamefont{Rotello}},
  \bibinfo{author}{\bibfnamefont{C.~C.} \bibnamefont{Kao}}, \bibnamefont{and}
  \bibinfo{author}{\bibfnamefont{B.}~\bibnamefont{Dragnea}},
  \bibinfo{journal}{Nano Letters} \textbf{\bibinfo{volume}{6}},
  \bibinfo{pages}{1993} (\bibinfo{year}{2006}).

\bibitem[{\citenamefont{Chen et~al.}(2005)\citenamefont{Chen, Kwak, Stein, Kao,
  and Dragnea}}]{Chen2005}
\bibinfo{author}{\bibfnamefont{C.}~\bibnamefont{Chen}},
  \bibinfo{author}{\bibfnamefont{E.~S.} \bibnamefont{Kwak}},
  \bibinfo{author}{\bibfnamefont{B.}~\bibnamefont{Stein}},
  \bibinfo{author}{\bibfnamefont{C.~C.} \bibnamefont{Kao}}, \bibnamefont{and}
  \bibinfo{author}{\bibfnamefont{B.}~\bibnamefont{Dragnea}},
  \bibinfo{journal}{J. Nanosci. and Nanotech.} \textbf{\bibinfo{volume}{5}},
  \bibinfo{pages}{2029} (\bibinfo{year}{2005}).

\bibitem[{\citenamefont{Dragnea et~al.}(2003)\citenamefont{Dragnea, Chen, Kwak,
  Stein, and Kao}}]{Dragnea2003}
\bibinfo{author}{\bibfnamefont{B.}~\bibnamefont{Dragnea}},
  \bibinfo{author}{\bibfnamefont{C.}~\bibnamefont{Chen}},
  \bibinfo{author}{\bibfnamefont{E.~S.} \bibnamefont{Kwak}},
  \bibinfo{author}{\bibfnamefont{B.}~\bibnamefont{Stein}}, \bibnamefont{and}
  \bibinfo{author}{\bibfnamefont{C.~C.} \bibnamefont{Kao}},
  \bibinfo{journal}{J. Am. Chem. Soc.} \textbf{\bibinfo{volume}{125}},
  \bibinfo{pages}{6374} (\bibinfo{year}{2003}).

\bibitem[{\citenamefont{Soto et~al.}(2006)\citenamefont{Soto, Blum, Vora,
  Lebedev, Meador, Won, Chatterji, Johnson, and Ratna}}]{Soto2006}
\bibinfo{author}{\bibfnamefont{C.~M.} \bibnamefont{Soto}},
  \bibinfo{author}{\bibfnamefont{A.~S.} \bibnamefont{Blum}},
  \bibinfo{author}{\bibfnamefont{G.~J.} \bibnamefont{Vora}},
  \bibinfo{author}{\bibfnamefont{N.}~\bibnamefont{Lebedev}},
  \bibinfo{author}{\bibfnamefont{C.~E.} \bibnamefont{Meador}},
  \bibinfo{author}{\bibfnamefont{A.~P.} \bibnamefont{Won}},
  \bibinfo{author}{\bibfnamefont{A.}~\bibnamefont{Chatterji}},
  \bibinfo{author}{\bibfnamefont{J.~E.} \bibnamefont{Johnson}},
  \bibnamefont{and} \bibinfo{author}{\bibfnamefont{B.~R.} \bibnamefont{Ratna}},
  \bibinfo{journal}{J. Am. Chem. Soc.} \textbf{\bibinfo{volume}{128}},
  \bibinfo{pages}{5184} (\bibinfo{year}{2006}).

\bibitem[{\citenamefont{Sapsford et~al.}(2006)\citenamefont{Sapsford, Soto,
  Blum, Chatterji, Lin, Johnson, Ligler, and Ratna}}]{Sapsford2006}
\bibinfo{author}{\bibfnamefont{K.~E.} \bibnamefont{Sapsford}},
  \bibinfo{author}{\bibfnamefont{C.~M.} \bibnamefont{Soto}},
  \bibinfo{author}{\bibfnamefont{A.~S.} \bibnamefont{Blum}},
  \bibinfo{author}{\bibfnamefont{A.}~\bibnamefont{Chatterji}},
  \bibinfo{author}{\bibfnamefont{T.~W.} \bibnamefont{Lin}},
  \bibinfo{author}{\bibfnamefont{J.~E.} \bibnamefont{Johnson}},
  \bibinfo{author}{\bibfnamefont{F.~S.} \bibnamefont{Ligler}},
  \bibnamefont{and} \bibinfo{author}{\bibfnamefont{B.~R.} \bibnamefont{Ratna}},
  \bibinfo{journal}{Biosens. Bioelectron.} \textbf{\bibinfo{volume}{21}},
  \bibinfo{pages}{1668} (\bibinfo{year}{2006}).

\bibitem[{\citenamefont{Boldogkoi et~al.}(2004)\citenamefont{Boldogkoi, Sik,
  Denes, Reichart, Toldi, Gerendai, Kovacs, and Palkovits}}]{Boldogkoi2004}
\bibinfo{author}{\bibfnamefont{Z.}~\bibnamefont{Boldogkoi}},
  \bibinfo{author}{\bibfnamefont{A.}~\bibnamefont{Sik}},
  \bibinfo{author}{\bibfnamefont{A.}~\bibnamefont{Denes}},
  \bibinfo{author}{\bibfnamefont{A.}~\bibnamefont{Reichart}},
  \bibinfo{author}{\bibfnamefont{J.}~\bibnamefont{Toldi}},
  \bibinfo{author}{\bibfnamefont{I.}~\bibnamefont{Gerendai}},
  \bibinfo{author}{\bibfnamefont{K.~J.} \bibnamefont{Kovacs}},
  \bibnamefont{and}
  \bibinfo{author}{\bibfnamefont{M.}~\bibnamefont{Palkovits}},
  \bibinfo{journal}{Prog. Neurobiol.} \textbf{\bibinfo{volume}{72}},
  \bibinfo{pages}{417} (\bibinfo{year}{2004}).

\bibitem[{\citenamefont{Gupta et~al.}(2005)\citenamefont{Gupta, Levchenko, and
  Torchilin}}]{Gupta2005}
\bibinfo{author}{\bibfnamefont{B.}~\bibnamefont{Gupta}},
  \bibinfo{author}{\bibfnamefont{T.~S.} \bibnamefont{Levchenko}},
  \bibnamefont{and} \bibinfo{author}{\bibfnamefont{V.~P.}
  \bibnamefont{Torchilin}}, \bibinfo{journal}{Advanced Drug Delivery Reviews}
  \textbf{\bibinfo{volume}{57}}, \bibinfo{pages}{637} (\bibinfo{year}{2005}).

\bibitem[{\citenamefont{Garcea and Gissmann}(2004)}]{Garcea2004}
\bibinfo{author}{\bibfnamefont{R.~L.} \bibnamefont{Garcea}} \bibnamefont{and}
  \bibinfo{author}{\bibfnamefont{L.}~\bibnamefont{Gissmann}},
  \bibinfo{journal}{Curr. Opin. Biotechnol.} \textbf{\bibinfo{volume}{15}},
  \bibinfo{pages}{513} (\bibinfo{year}{2004}).

\bibitem[{\citenamefont{Dietz and Bahr}(2004)}]{Dietz2004}
\bibinfo{author}{\bibfnamefont{G.~P.~H.} \bibnamefont{Dietz}} \bibnamefont{and}
  \bibinfo{author}{\bibfnamefont{M.}~\bibnamefont{Bahr}},
  \bibinfo{journal}{Molecular and Cellular Neuroscience}
  \textbf{\bibinfo{volume}{27}}, \bibinfo{pages}{85} (\bibinfo{year}{2004}).

\bibitem[{\citenamefont{Chatterji et~al.}(2005)\citenamefont{Chatterji, Ochoa,
  Ueno, Lin, and Johnson}}]{Chatterji2005}
\bibinfo{author}{\bibfnamefont{A.}~\bibnamefont{Chatterji}},
  \bibinfo{author}{\bibfnamefont{W.~F.} \bibnamefont{Ochoa}},
  \bibinfo{author}{\bibfnamefont{T.}~\bibnamefont{Ueno}},
  \bibinfo{author}{\bibfnamefont{T.~W.} \bibnamefont{Lin}}, \bibnamefont{and}
  \bibinfo{author}{\bibfnamefont{J.~E.} \bibnamefont{Johnson}},
  \bibinfo{journal}{Nano Letters} \textbf{\bibinfo{volume}{5}},
  \bibinfo{pages}{597} (\bibinfo{year}{2005}).

\bibitem[{\citenamefont{Falkner et~al.}(2005)\citenamefont{Falkner, Turner,
  Bosworth, Trentler, Johnson, Lin, and Colvin}}]{Falkner2005}
\bibinfo{author}{\bibfnamefont{J.~C.} \bibnamefont{Falkner}},
  \bibinfo{author}{\bibfnamefont{M.~E.} \bibnamefont{Turner}},
  \bibinfo{author}{\bibfnamefont{J.~K.} \bibnamefont{Bosworth}},
  \bibinfo{author}{\bibfnamefont{T.~J.} \bibnamefont{Trentler}},
  \bibinfo{author}{\bibfnamefont{J.~E.} \bibnamefont{Johnson}},
  \bibinfo{author}{\bibfnamefont{T.~W.} \bibnamefont{Lin}}, \bibnamefont{and}
  \bibinfo{author}{\bibfnamefont{V.~L.} \bibnamefont{Colvin}},
  \bibinfo{journal}{J. Am. Chem. Soc.} \textbf{\bibinfo{volume}{127}},
  \bibinfo{pages}{5274} (\bibinfo{year}{2005}).

\bibitem[{\citenamefont{Flynn et~al.}(2003)\citenamefont{Flynn, Lee, Peelle,
  and Belcher}}]{Flynn2003}
\bibinfo{author}{\bibfnamefont{C.~E.} \bibnamefont{Flynn}},
  \bibinfo{author}{\bibfnamefont{S.~W.} \bibnamefont{Lee}},
  \bibinfo{author}{\bibfnamefont{B.~R.} \bibnamefont{Peelle}},
  \bibnamefont{and} \bibinfo{author}{\bibfnamefont{A.~M.}
  \bibnamefont{Belcher}}, \bibinfo{journal}{Acta Materialia}
  \textbf{\bibinfo{volume}{51}}, \bibinfo{pages}{5867} (\bibinfo{year}{2003}).

\bibitem[{\citenamefont{Douglas and Young}(1998)}]{Douglas1998}
\bibinfo{author}{\bibfnamefont{T.}~\bibnamefont{Douglas}} \bibnamefont{and}
  \bibinfo{author}{\bibfnamefont{M.}~\bibnamefont{Young}},
  \bibinfo{journal}{Nature} \textbf{\bibinfo{volume}{393}},
  \bibinfo{pages}{152} (\bibinfo{year}{1998}).

\bibitem[{\citenamefont{Ceres and Zlotnick}(2002)}]{Ceres2002}
\bibinfo{author}{\bibfnamefont{P.}~\bibnamefont{Ceres}} \bibnamefont{and}
  \bibinfo{author}{\bibfnamefont{A.}~\bibnamefont{Zlotnick}},
  \bibinfo{journal}{Biochemistry} \textbf{\bibinfo{volume}{41}},
  \bibinfo{pages}{11525} (\bibinfo{year}{2002}).

\bibitem[{\citenamefont{Endres and Zlotnick}(2002)}]{Endres2002}
\bibinfo{author}{\bibfnamefont{D.}~\bibnamefont{Endres}} \bibnamefont{and}
  \bibinfo{author}{\bibfnamefont{A.}~\bibnamefont{Zlotnick}},
  \bibinfo{journal}{Biophys. J.} \textbf{\bibinfo{volume}{83}},
  \bibinfo{pages}{1217} (\bibinfo{year}{2002}).

\bibitem[{\citenamefont{Zhang and Schwartz}(2006)}]{Zhang2006}
\bibinfo{author}{\bibfnamefont{T.~Q.} \bibnamefont{Zhang}} \bibnamefont{and}
  \bibinfo{author}{\bibfnamefont{R.}~\bibnamefont{Schwartz}},
  \bibinfo{journal}{Biophys. J.} \textbf{\bibinfo{volume}{90}},
  \bibinfo{pages}{57} (\bibinfo{year}{2006}).

\bibitem[{\citenamefont{Zlotnick et~al.}(1999)\citenamefont{Zlotnick, Johnson,
  Wingfield, Stahl, and Endres}}]{Zlotnick1999}
\bibinfo{author}{\bibfnamefont{A.}~\bibnamefont{Zlotnick}},
  \bibinfo{author}{\bibfnamefont{J.~M.} \bibnamefont{Johnson}},
  \bibinfo{author}{\bibfnamefont{P.~W.} \bibnamefont{Wingfield}},
  \bibinfo{author}{\bibfnamefont{S.~J.} \bibnamefont{Stahl}}, \bibnamefont{and}
  \bibinfo{author}{\bibfnamefont{D.}~\bibnamefont{Endres}},
  \bibinfo{journal}{Biochemistry} \textbf{\bibinfo{volume}{38}},
  \bibinfo{pages}{14644} (\bibinfo{year}{1999}).

\bibitem[{\citenamefont{Sorger et~al.}(1986)\citenamefont{Sorger, Stockley, and
  Harrison}}]{Sorger1986}
\bibinfo{author}{\bibfnamefont{P.~K.} \bibnamefont{Sorger}},
  \bibinfo{author}{\bibfnamefont{P.~G.} \bibnamefont{Stockley}},
  \bibnamefont{and} \bibinfo{author}{\bibfnamefont{S.~C.}
  \bibnamefont{Harrison}}, \bibinfo{journal}{J. Mol. Biol.}
  \textbf{\bibinfo{volume}{191}}, \bibinfo{pages}{639} (\bibinfo{year}{1986}).

\bibitem[{\citenamefont{Schwartz et~al.}(1998)\citenamefont{Schwartz, Shor,
  Prevelige, and Berger}}]{Schwartz1998}
\bibinfo{author}{\bibfnamefont{R.}~\bibnamefont{Schwartz}},
  \bibinfo{author}{\bibfnamefont{P.~W.} \bibnamefont{Shor}},
  \bibinfo{author}{\bibfnamefont{P.~E.} \bibnamefont{Prevelige}},
  \bibnamefont{and} \bibinfo{author}{\bibfnamefont{B.}~\bibnamefont{Berger}},
  \bibinfo{journal}{Biophys. J.} \textbf{\bibinfo{volume}{75}},
  \bibinfo{pages}{2626} (\bibinfo{year}{1998}).

\bibitem[{\citenamefont{Jack et~al.}(2007)\citenamefont{Jack, Hagan, and
  Chandler}}]{Jack2007}
\bibinfo{author}{\bibfnamefont{R.~L.} \bibnamefont{Jack}},
  \bibinfo{author}{\bibfnamefont{M.~F.} \bibnamefont{Hagan}}, \bibnamefont{and}
  \bibinfo{author}{\bibfnamefont{D.}~\bibnamefont{Chandler}},
  \bibinfo{journal}{Phys. Rev. E} \textbf{\bibinfo{volume}{76}},
  \bibinfo{pages}{021119} (\bibinfo{year}{2007}).

\bibitem[{\citenamefont{Whitesides and Boncheva}(2002)}]{Whitesides2002}
\bibinfo{author}{\bibfnamefont{G.~M.} \bibnamefont{Whitesides}}
  \bibnamefont{and} \bibinfo{author}{\bibfnamefont{M.}~\bibnamefont{Boncheva}},
  \bibinfo{journal}{Proc. Natl. Acad. Sci. U. S. A.}
  \textbf{\bibinfo{volume}{99}}, \bibinfo{pages}{4769} (\bibinfo{year}{2002}).

\bibitem[{\citenamefont{Johnson et~al.}(2004)\citenamefont{Johnson, Willits,
  Young, and Zlotnick}}]{Johnson2004}
\bibinfo{author}{\bibfnamefont{J.~M.} \bibnamefont{Johnson}},
  \bibinfo{author}{\bibfnamefont{D.~A.} \bibnamefont{Willits}},
  \bibinfo{author}{\bibfnamefont{M.~J.} \bibnamefont{Young}}, \bibnamefont{and}
  \bibinfo{author}{\bibfnamefont{A.}~\bibnamefont{Zlotnick}},
  \bibinfo{journal}{J. Mol. Biol.} \textbf{\bibinfo{volume}{335}},
  \bibinfo{pages}{455} (\bibinfo{year}{2004}).

\bibitem[{\citenamefont{Angelescu et~al.}(2006)\citenamefont{Angelescu,
  Bruinsma, and Linse}}]{Angelescu2006}
\bibinfo{author}{\bibfnamefont{D.~G.} \bibnamefont{Angelescu}},
  \bibinfo{author}{\bibfnamefont{R.}~\bibnamefont{Bruinsma}}, \bibnamefont{and}
  \bibinfo{author}{\bibfnamefont{P.}~\bibnamefont{Linse}},
  \bibinfo{journal}{Phys. Rev. E} \textbf{\bibinfo{volume}{73}},
  \bibinfo{pages}{041921} (\bibinfo{year}{2006}).

\bibitem[{\citenamefont{van~der Schoot and Bruinsma}(2005)}]{vanderSchoot2005}
\bibinfo{author}{\bibfnamefont{P.}~\bibnamefont{van~der Schoot}}
  \bibnamefont{and} \bibinfo{author}{\bibfnamefont{R.}~\bibnamefont{Bruinsma}},
  \bibinfo{journal}{Phys. Rev. E} \textbf{\bibinfo{volume}{71}},
  \bibinfo{pages}{061928} (\bibinfo{year}{2005}).

\bibitem[{\citenamefont{Zhang et~al.}(2004)\citenamefont{Zhang, Konecny, Baker,
  and McCammon}}]{Zhang2004}
\bibinfo{author}{\bibfnamefont{D.~Q.} \bibnamefont{Zhang}},
  \bibinfo{author}{\bibfnamefont{R.}~\bibnamefont{Konecny}},
  \bibinfo{author}{\bibfnamefont{N.~A.} \bibnamefont{Baker}}, \bibnamefont{and}
  \bibinfo{author}{\bibfnamefont{J.~A.} \bibnamefont{McCammon}},
  \bibinfo{journal}{Biopolymers} \textbf{\bibinfo{volume}{75}},
  \bibinfo{pages}{325} (\bibinfo{year}{2004}).

\bibitem[{\citenamefont{Hu et~al.}(2007)\citenamefont{Hu, Zhang, and
  Shklovskii}}]{Hu2007b}
\bibinfo{author}{\bibfnamefont{T.}~\bibnamefont{Hu}},
  \bibinfo{author}{\bibfnamefont{R.}~\bibnamefont{Zhang}}, \bibnamefont{and}
  \bibinfo{author}{\bibfnamefont{B.~I.} \bibnamefont{Shklovskii}},
  \bibinfo{journal}{arXiv:q-bio/0610009v4}  (\bibinfo{year}{2007}).

\bibitem[{\citenamefont{Chen et~al.}(2007)\citenamefont{Chen, Zhang, and
  Glotzer}}]{Chen2007b}
\bibinfo{author}{\bibfnamefont{T.}~\bibnamefont{Chen}},
  \bibinfo{author}{\bibfnamefont{Z.~L.} \bibnamefont{Zhang}}, \bibnamefont{and}
  \bibinfo{author}{\bibfnamefont{S.~C.} \bibnamefont{Glotzer}},
  \bibinfo{journal}{Proc. Natl. Acad. Sci. U. S. A.}
  \textbf{\bibinfo{volume}{104}}, \bibinfo{pages}{717} (\bibinfo{year}{2007}).

\bibitem[{\citenamefont{Rudnick and Bruinsma}(2005)}]{Rudnick2005}
\bibinfo{author}{\bibfnamefont{J.}~\bibnamefont{Rudnick}} \bibnamefont{and}
  \bibinfo{author}{\bibfnamefont{R.}~\bibnamefont{Bruinsma}},
  \bibinfo{journal}{Phys. Rev. Lett.} \textbf{\bibinfo{volume}{94}},
  \bibinfo{pages}{038101} (\bibinfo{year}{2005}).

\bibitem[{\citenamefont{Hu and Shklovskii}(2007)}]{Hu2007}
\bibinfo{author}{\bibfnamefont{T.}~\bibnamefont{Hu}} \bibnamefont{and}
  \bibinfo{author}{\bibfnamefont{B.~I.} \bibnamefont{Shklovskii}},
  \bibinfo{journal}{Phys. Rev. E} \textbf{\bibinfo{volume}{75}},
  \bibinfo{pages}{051901} (\bibinfo{year}{2007}).

\bibitem[{\citenamefont{McPherson}(2005)}]{McPherson2005}
\bibinfo{author}{\bibfnamefont{A.}~\bibnamefont{McPherson}},
  \bibinfo{journal}{BioEssays} \textbf{\bibinfo{volume}{27}},
  \bibinfo{pages}{447} (\bibinfo{year}{2005}).

\bibitem[{\citenamefont{Dragnea}(2007)}]{Dragnea2007}
\bibinfo{author}{\bibfnamefont{B.}~\bibnamefont{Dragnea}},
  \emph{\bibinfo{title}{Personal communication}} (\bibinfo{year}{2007}).

\bibitem[{\citenamefont{van Blaaderen}(2006)}]{vanBlaaderen2006}
\bibinfo{author}{\bibfnamefont{A.}~\bibnamefont{van Blaaderen}},
  \bibinfo{journal}{Nature} \textbf{\bibinfo{volume}{439}},
  \bibinfo{pages}{545} (\bibinfo{year}{2006}).

\bibitem[{\citenamefont{Li et~al.}(2005)\citenamefont{Li, Lee, Rubner, and
  Cohen}}]{Li2005}
\bibinfo{author}{\bibfnamefont{Z.~F.} \bibnamefont{Li}},
  \bibinfo{author}{\bibfnamefont{D.~Y.} \bibnamefont{Lee}},
  \bibinfo{author}{\bibfnamefont{M.~F.} \bibnamefont{Rubner}},
  \bibnamefont{and} \bibinfo{author}{\bibfnamefont{R.~E.} \bibnamefont{Cohen}},
  \bibinfo{journal}{Macromolecules} \textbf{\bibinfo{volume}{38}},
  \bibinfo{pages}{7876} (\bibinfo{year}{2005}).

\bibitem[{\citenamefont{Cho et~al.}(2005)\citenamefont{Cho, Yi, Lim, Kim,
  Manoharan, Pine, and Yang}}]{Cho2005}
\bibinfo{author}{\bibfnamefont{Y.~S.} \bibnamefont{Cho}},
  \bibinfo{author}{\bibfnamefont{G.~R.} \bibnamefont{Yi}},
  \bibinfo{author}{\bibfnamefont{J.~M.} \bibnamefont{Lim}},
  \bibinfo{author}{\bibfnamefont{S.~H.} \bibnamefont{Kim}},
  \bibinfo{author}{\bibfnamefont{V.~N.} \bibnamefont{Manoharan}},
  \bibinfo{author}{\bibfnamefont{D.~J.} \bibnamefont{Pine}}, \bibnamefont{and}
  \bibinfo{author}{\bibfnamefont{S.~M.} \bibnamefont{Yang}},
  \bibinfo{journal}{J. Am. Chem. Soc.} \textbf{\bibinfo{volume}{127}},
  \bibinfo{pages}{15968} (\bibinfo{year}{2005}).

\bibitem[{\citenamefont{Speir et~al.}(1995)\citenamefont{Speir, Munshi, Wang,
  Baker, and Johnson}}]{Speir1995}
\bibinfo{author}{\bibfnamefont{J.~A.} \bibnamefont{Speir}},
  \bibinfo{author}{\bibfnamefont{S.}~\bibnamefont{Munshi}},
  \bibinfo{author}{\bibfnamefont{G.~J.} \bibnamefont{Wang}},
  \bibinfo{author}{\bibfnamefont{T.~S.} \bibnamefont{Baker}}, \bibnamefont{and}
  \bibinfo{author}{\bibfnamefont{J.~E.} \bibnamefont{Johnson}},
  \bibinfo{journal}{Structure} \textbf{\bibinfo{volume}{3}},
  \bibinfo{pages}{63} (\bibinfo{year}{1995}).

\bibitem[{\citenamefont{Parent et~al.}(2007)\citenamefont{Parent, Suhanovsky,
  and Teschke}}]{Parent2007}
\bibinfo{author}{\bibfnamefont{K.~N.} \bibnamefont{Parent}},
  \bibinfo{author}{\bibfnamefont{M.~M.} \bibnamefont{Suhanovsky}},
  \bibnamefont{and} \bibinfo{author}{\bibfnamefont{C.~M.}
  \bibnamefont{Teschke}}, \bibinfo{journal}{J. Mol. Biol.}
  \textbf{\bibinfo{volume}{365}}, \bibinfo{pages}{513} (\bibinfo{year}{2007}).

\bibitem[{\citenamefont{Prevelige et~al.}(1993)\citenamefont{Prevelige, Thomas,
  and King}}]{Prevelige1993}
\bibinfo{author}{\bibfnamefont{P.~E.} \bibnamefont{Prevelige}},
  \bibinfo{author}{\bibfnamefont{D.}~\bibnamefont{Thomas}}, \bibnamefont{and}
  \bibinfo{author}{\bibfnamefont{J.}~\bibnamefont{King}},
  \bibinfo{journal}{Biophys. J.} \textbf{\bibinfo{volume}{64}},
  \bibinfo{pages}{824} (\bibinfo{year}{1993}).

\bibitem[{\citenamefont{Lucas et~al.}(2002)\citenamefont{Lucas, Larson, and
  McPherson}}]{Lucas2002}
\bibinfo{author}{\bibfnamefont{R.~W.} \bibnamefont{Lucas}},
  \bibinfo{author}{\bibfnamefont{S.~B.} \bibnamefont{Larson}},
  \bibnamefont{and}
  \bibinfo{author}{\bibfnamefont{A.}~\bibnamefont{McPherson}},
  \bibinfo{journal}{J. Mol. Biol.} \textbf{\bibinfo{volume}{317}},
  \bibinfo{pages}{95} (\bibinfo{year}{2002}).

\bibitem[{\citenamefont{Natarajan et~al.}(2005)\citenamefont{Natarajan, Lander,
  Shepherd, Reddy, Brooks, and Johnson}}]{Natarajan2005}
\bibinfo{author}{\bibfnamefont{P.}~\bibnamefont{Natarajan}},
  \bibinfo{author}{\bibfnamefont{G.~C.} \bibnamefont{Lander}},
  \bibinfo{author}{\bibfnamefont{C.~M.} \bibnamefont{Shepherd}},
  \bibinfo{author}{\bibfnamefont{V.~S.} \bibnamefont{Reddy}},
  \bibinfo{author}{\bibfnamefont{C.~L.} \bibnamefont{Brooks}},
  \bibnamefont{and} \bibinfo{author}{\bibfnamefont{J.~E.}
  \bibnamefont{Johnson}}, \bibinfo{journal}{Nature Reviews Microbiology}
  \textbf{\bibinfo{volume}{3}}, \bibinfo{pages}{809} (\bibinfo{year}{2005}).

\bibitem[{\citenamefont{Frenkel and Smit}(2002)}]{Frenkel2002}
\bibinfo{author}{\bibfnamefont{D.}~\bibnamefont{Frenkel}} \bibnamefont{and}
  \bibinfo{author}{\bibfnamefont{B.}~\bibnamefont{Smit}},
  \emph{\bibinfo{title}{Understanding molecular simulation: from algorithms to
  applications}} (\bibinfo{publisher}{Academic}, \bibinfo{address}{San Diego,
  Calif. ; London}, \bibinfo{year}{2002}), \bibinfo{edition}{2nd} ed.

\bibitem[{\citenamefont{Kamber et~al.}(2007)\citenamefont{Kamber, Kondev, and
  Hagan}}]{Kamber2007}
\bibinfo{author}{\bibfnamefont{E.}~\bibnamefont{Kamber}},
  \bibinfo{author}{\bibfnamefont{J.}~\bibnamefont{Kondev}}, \bibnamefont{and}
  \bibinfo{author}{\bibfnamefont{M.~F.} \bibnamefont{Hagan}},
  \bibinfo{journal}{in preparation}  (\bibinfo{year}{2007}).

\bibitem[{\citenamefont{Wilber et~al.}(2007)\citenamefont{Wilber, Doye, Louis,
  Noya, Miller, and Wong}}]{Wilber2007}
\bibinfo{author}{\bibfnamefont{A.~W.} \bibnamefont{Wilber}},
  \bibinfo{author}{\bibfnamefont{J.~P.~K.} \bibnamefont{Doye}},
  \bibinfo{author}{\bibfnamefont{A.~A.} \bibnamefont{Louis}},
  \bibinfo{author}{\bibfnamefont{E.~G.} \bibnamefont{Noya}},
  \bibinfo{author}{\bibfnamefont{M.~A.} \bibnamefont{Miller}},
  \bibnamefont{and} \bibinfo{author}{\bibfnamefont{P.}~\bibnamefont{Wong}},
  \bibinfo{journal}{J. Chem. Phys.} \textbf{\bibinfo{volume}{127}}
  (\bibinfo{year}{2007}).

\bibitem[{\citenamefont{Zandi et~al.}(2006)\citenamefont{Zandi, van~der Schoot,
  Reguera, Kegel, and Reiss}}]{Zandi2006}
\bibinfo{author}{\bibfnamefont{R.}~\bibnamefont{Zandi}},
  \bibinfo{author}{\bibfnamefont{P.}~\bibnamefont{van~der Schoot}},
  \bibinfo{author}{\bibfnamefont{D.}~\bibnamefont{Reguera}},
  \bibinfo{author}{\bibfnamefont{W.}~\bibnamefont{Kegel}}, \bibnamefont{and}
  \bibinfo{author}{\bibfnamefont{H.}~\bibnamefont{Reiss}},
  \bibinfo{journal}{Biophys. J.} \textbf{\bibinfo{volume}{90}},
  \bibinfo{pages}{1939} (\bibinfo{year}{2006}).

\bibitem[{\citenamefont{Berg and Purcell}(1977)}]{Berg1977}
\bibinfo{author}{\bibfnamefont{H.~C.} \bibnamefont{Berg}} \bibnamefont{and}
  \bibinfo{author}{\bibfnamefont{E.~M.} \bibnamefont{Purcell}},
  \bibinfo{journal}{Biophys. J.} \textbf{\bibinfo{volume}{20}},
  \bibinfo{pages}{193} (\bibinfo{year}{1977}).

\bibitem[{\citenamefont{Humphrey et~al.}(1996)\citenamefont{Humphrey, Dalke,
  and Schulten}}]{Humphrey1996}
\bibinfo{author}{\bibfnamefont{W.}~\bibnamefont{Humphrey}},
  \bibinfo{author}{\bibfnamefont{A.}~\bibnamefont{Dalke}}, \bibnamefont{and}
  \bibinfo{author}{\bibfnamefont{K.}~\bibnamefont{Schulten}},
  \bibinfo{journal}{J. Mol. Graph.} \textbf{\bibinfo{volume}{14}},
  \bibinfo{pages}{33} (\bibinfo{year}{1996}).

\bibitem[{\citenamefont{Krol et~al.}(1999)\citenamefont{Krol, Olson, Tate,
  Johnson, Baker, and Ahlquist}}]{Krol1999}
\bibinfo{author}{\bibfnamefont{M.~A.} \bibnamefont{Krol}},
  \bibinfo{author}{\bibfnamefont{N.~H.} \bibnamefont{Olson}},
  \bibinfo{author}{\bibfnamefont{J.}~\bibnamefont{Tate}},
  \bibinfo{author}{\bibfnamefont{J.~E.} \bibnamefont{Johnson}},
  \bibinfo{author}{\bibfnamefont{T.~S.} \bibnamefont{Baker}}, \bibnamefont{and}
  \bibinfo{author}{\bibfnamefont{P.}~\bibnamefont{Ahlquist}},
  \bibinfo{journal}{Proc. Natl. Acad. Sci. U. S. A.}
  \textbf{\bibinfo{volume}{96}}, \bibinfo{pages}{13650} (\bibinfo{year}{1999}).

\end{thebibliography}

\end{document}